\documentclass[12pt]{article}
\usepackage[T1]{fontenc}
\usepackage{tikz}
\usetikzlibrary{automata, positioning, arrows, shapes}

\usepackage{pgfplots}
\usepgfplotslibrary{groupplots}
\usetikzlibrary{pgfplots.groupplots}

\usepackage{caption}
\usepackage{subcaption}
\usepackage{setspace,graphicx,epstopdf,amsmath,amsfonts,amssymb,amsthm,versionPO}
\usepackage{marginnote,datetime,enumitem,subfig,rotating,fancyvrb}

\usepackage[longnamesfirst]{natbib}

\newcommand{\mathbbm}[1]{\text{\usefont{U}{bbm}{m}{n}#1}}

\usdate

\excludeversion{notes}
\includeversion{links}

\iflinks{}{\hypersetup{draft=true}}

\usepackage[margin=1in]{geometry}

\makeatletter\let\chapter\@undefined\makeatother

\usepackage{indentfirst}
\usepackage{endnotes}
\usepackage{jf}
\usepackage[labelfont=bf,labelsep=period]{caption}
\usepackage{hyperref,float}

\newtheorem{corollary}{COROLLARY}
\newtheorem{proposition}{PROPOSITION}

\newtheorem{lem}{LEMMA}
\newtheorem{assume}{ASSUMPTION}
\newcommand{\argmax}{\mathop{\rm arg\,max}}

\makeatletter
\providecommand\sf@counterlist{}
\makeatother

\begin{document}

\setlist{noitemsep}
\onehalfspacing

\author{Agostino Capponi\thanks{\rm A. Capponi is with the Industrial Engineering and Operations Research, Columbia University.} \and Álvaro Cartea\thanks{\rm Á. Cartea is with the Oxford-Man Institute and the Mathematical Institute, University of Oxford.} \and Fayçal Drissi\thanks{\rm F. Drissi is with the Oxford-Man Institute, University of Oxford. \\\textbf{Acknowledgments}: We are grateful to Bruno Biais, Patrick Chang, Zachary Feinstein, Joel Hasbrouck, Sebastian Jaimungal, Olga Klein, Igor Makarov, Katya Malinova, Andreas Park, Fahad Saleh, and Basil Williams for their valuable comments. We are grateful to the organizers of the NBER Summer Institute 2026, Financial Market Structure, and the 41st Meeting of the European Economic Association and 77th European Meeting of the Econometric Society (EEA-ESEM 2026) for accepting the paper in their programs. We are also thankful to participants of the CBER Conference at Columbia University, the Oxford-Harvard Conference on Decentralised Finance  and Market Microstructure, the Stevens Institute of Technology's DeFi Seminar, the SEEM Seminar at the Chinese University of Hong Kong, the 2025 Warwick Business School Workshop on Financial Technology, the session on Blockchain Economics and Decentralized Finance of the INFORMS $2025$ annual meeting, and the Oxford-Princeton Workshop  for insightful discussions.}}

\title{
\bf
The Viability of Blockchain Markets under Discrete Clearing and Paid Priority
}

\date{Latest \href{https://fdr0903.github.io/files/mempools.pdf}{version}. \\ This version: \today. \\ First version: March 21, 2025.}

\maketitle
\thispagestyle{empty}

\centerline{\bf ABSTRACT}

\begin{doublespace}
\noindent
\small

This paper develops a model to evaluate the viability of blockchain markets as the sole venue for price formation. Blockchains clear at discrete intervals called \textit{block time}, and transactions are executed sequentially according to \textit{priority fees} paid by traders who compete for queue position.  We show that these features undermine the viability of markets. Paid-priority ordering induces endogenous selection, where only traders with sufficiently high valuations participate. The participation cutoff rises with competition, which intensifies with lower information costs or higher liquidity demand. This hinders price discovery and biases prices. It also impairs liquidity: the cutoff concentrates trading among aggressive traders and increases adverse selection that liquidity suppliers absorb in a single clearing round. Although longer block times enhance consensus security, they amplify these effects and can cause markets to shut down.

\end{doublespace}

\medskip

\begin{doublespace}
\clearpage

A central ambition of blockchain technology is to provide a decentralized infrastructure to organize economic activity at scale. The viability of financial markets within this infrastructure is therefore a fundamental test of that ambition. This paper develops a theoretical framework that models the blockchain as the sole venue for price formation to assess its viability as an infrastructure to support markets. The model endogenizes liquidity supply, trading volumes, and information revealed in equilibrium. It incorporates three defining features of blockchain infrastructure: clearing occurs discretely in a single round within each block; block time determines the length of the round and is necessary to conduct security consensus and preserve decentralization; transactions are ordered through paid-priority discriminatory pricing.

Blockchains operate in discrete time, and information is revealed through an auction with discriminatory pricing. Orders are recorded into \emph{blocks} built at regular intervals, called \emph{slots}, and the duration of a slot is referred to as \emph{block time}. At the end of a slot, a \textit{builder} is selected by the blockchain's \emph{protocol} to build the next block. Once the block is created, the transactions composing the block are executed in order of inclusion. The builder sequences transactions in the block according to \emph{priority fees} paid by traders who compete for queue position. Consequently, traders engage in \emph{priority gas auctions} (PGAs) to optimize their queue positions in the block. Blockchain trading contrasts with traditional markets with continuous trading. In the latter, trading unfolds over multiple short clearing rounds, and in the former, trading unfolds over relatively long clearing rounds with discriminatory prices.

We model the strategic interaction between liquidity suppliers and liquidity takers for a risky asset on a blockchain in three stages. In stage one, traders decide whether to enter the blockchain and acquire information at a cost. In stage two, a liquidity supplier sets the depth of a decentralized exchange (DEX) by balancing revenue from price-sensitive uninformed demand against expected losses to informed traders.\footnote{The DEX operates as a smart contract under the blockchain protocol. DEXs are the most widely used type of blockchain market and employ algorithms to price and trade digital assets. In 2025, DEXs support monthly trading volumes of approximately~$\$420$~billion. A growing literature studies their microstructure (\cite{lehar2025decentralized, capponi2021adoption, barbon2021quality, john2023smart, klein2023price}) and compares them with traditional exchanges; see~\cite{barbon2021quality, aoyagi2021coexisting}. The blockchain protocol defines the rules governing how transactions are recorded and determines the lifecycle of orders.} In stage three, informed traders with private 
valuations compete in the PGA to buy or to sell the asset, and price-sensitive uninformed traders submit transactions. In our model, informed traders are either all buyers or all sellers. Block time is long enough for substantial price movements to occur, so most private valuations observed at the end of block time take a definite sign. The remainder of the introduction considers buyers; the seller case  is symmetric.

Traders compete for block priority to avoid worse execution prices resulting from the adverse price impact of preceding orders in the block. 
We propose and solve a special form of multi-prize auction for the PGA  in which each trader's payoff depends on the rank of their bid relative to that of others. 
Specifically, the surplus from bidding a priority fee is the expected price impact that a trader avoids by improving queue position. In the perfect Bayesian equilibrium of the PGA, informed traders determine both their priority fees and their trading volumes. To determine priority fees,  they balance profits from better execution prices against the fee paid to the builder. To determine volumes, they balance increased expected profits from larger positions against slippage that increases with order size.

We show that the defining features of blockchains fundamentally determine market outcomes and undermine market viability. Our main findings are threefold. First, competition in paid-priority contests generates endogenous selection, where only traders with sufficiently high valuations participate, which hinders price discovery by truncating information and biasing prices towards extreme values. Second, this selection mechanism raises adverse selection costs for liquidity providers and decreases liquidity in equilibrium, potentially leading to a market shutdown. Third, the very feature that supports decentralization, namely block time for consensus, amplifies these effects and  tightens the constraints for viable markets. We elaborate on each of these findings below.

Our first two findings indicate that increased competition between informed traders in blockchains generates adverse effects for both liquidity and price discovery. First, we show that there exists a cutoff for private valuations below which traders choose not to participate in the PGA, so only aggressive traders with high valuations transact in the blockchain. The intuition is as follows. Low valuations correspond to low desired volumes and less aggressiveness in priority fee bidding. This results in worse queue positions within the block and, consequently, higher adverse price impact from traders with better positions in the block and larger volumes. For traders with sufficiently low valuations, the adverse price impact associated with high-valuation traders will discourage participation because expected wealth from trading is negative. Hence, only a small fraction of traders in the blockchain with high enough valuations participate in the PGA.

Two factors exacerbate these negative effects on markets. First, as the number of competitors increases, the participation cutoff rises, so increasingly higher valuations are required to participate in the PGA. Thus, competition limits the amount of information that is disseminated: it narrows the range of private valuations revealed to the market, and skews it towards extreme values. Second, competition also impairs liquidity. Although the fraction of informed traders submitting transactions becomes smaller with competition, trading concentrates among the most aggressive traders, and the resulting adverse selection costs per unit of liquidity supply eventually increase.

These findings contrast with market outcomes in traditional markets where trading unfolds over multiple short clearing rounds. There, greater competition among informed traders accelerates the incorporation of information in the early rounds, and market depth typically improves thereafter; see \cite{holden1992long}. The mechanism behind our result is fundamentally different. Blockchain block times are long, so each blockchain round of clearing aggregates the trading activity that, in traditional markets, would unfold over multiple successive rounds. Two consequences follow. First, the liquidity supplier must absorb the adverse-selection risk from aggregate informed flow in a single decision, rather than adjusting liquidity across successive rounds. Second, paid-priority  induces endogenous selection and only the most aggressive informed traders participate, pushing prices toward extreme values and increasing adverse-selection costs. As a result, competition decreases liquidity.\footnote{When information is long-lived over multiple blocks, priority fees reveal valuations within the first block. Thus, blending with noise traders over multiple blocks would require a zero priority fee. This can be shown to be dominated by any positive fee that improves queue position. Hence informed traders do not delay revelation across blocks.}

Our model shows that a larger mass of uninformed demand and lower information costs increase the equilibrium number of informed traders who enter the blockchain and participate in the PGA. This would improve market efficiency in traditional electronic markets. The effect is reversed under paid-priority pricing. In blockchains, an increased number of informed traders raises the participation cutoff, so information becomes more truncated, and price efficiency deteriorates.

Our third main finding is that block time creates a tradeoff between security and efficiency.  Longer blocks increase the dispersion of private valuations and intensify adverse selection, while simultaneously reducing the effective mass of price-sensitive uninformed demand. As a result, with many informed traders, liquidity decreases and there is a critical block time beyond which the market collapses.

Our model implies several stylized facts that we take to the data.
First, empirical evidence shows that per-transaction volumes
in blockchain markets are significantly higher than those in
electronic limit order books for similar assets. This provides empirical support to the endogenous selection mechanism, whereby mostly high-valuation traders participate. Second, our model predicts that both trading volumes and priority fees increase with private valuations, and that higher private valuations correspond to better positions in the block. Using historical data from the DEX Uniswap v3 on Ethereum, we document that transaction volumes decrease with queue position, providing empirical support for this prediction.

\paragraph{\textbf{Literature review.}} The tradeoff faced by informed traders in PGAs is analogous to that faced by political contestants or researchers in R\&D races; see \cite{lazear1981rank, hillman1989politically, barut1998symmetric}. In such contests,  the cost of the effort exerted by contestants corresponds to the priority fee paid by traders, and the payoff linked to achieving a particular rank in the race corresponds to the surplus in the PGA; see also \cite{klose2015all}. Multi-prize contests are typically modeled as independent private cost contests, where the cost of effort is private but the prize vector is known ex ante; see~\cite{moldovanu2001optimal}.  In contrast, in the PGA, the \textit{prize} from attaining a given position is not fixed but depends on the private valuations and trading volumes of competitors. In the PGA, the value of a given position corresponds to the adverse price impact that a trader avoids by securing that position within the block.

The model of this paper is related to the literature that studies the properties of price and liquidity. Our model is information based, however, it shares features of the multi-stage inventory-based models in  \cite{biais1993price, de2002risk}. We adapt these models to account for information dissemination,  DEX algorithmic pricing, and the blockchain protocol whereby traders compete for queue position instead of speed (\cite{o1998market}). Our model is also related to the theory on the costs of information processing such as that in \cite{grossman1980impossibility}, \cite{verrecchia1982information},  \cite{kyle1989informed}, and \cite{baldauf2020high}.

Some studies examine periodic uniform price auctions for market clearing (\cite{madhavan1992trading, budish2015high}), and they show that these auctions enhance price efficiency but impose higher information costs on traders. Instead, our model considers discriminatory auctions in which liquidity suppliers do not participate. 

Several studies assume an information monopolist (\cite{kyle:1985}). In blockchains, monopolistic information would drive priority fees to zero in equilibrium, which contradicts empirical evidence showing that priority fees increase with the informational content of transactions; see \cite{capponi2024price}. We therefore consider competition among multiple informed traders. Models with competition, such as \cite{holden1992long} and \cite{foster1996strategic}, lead to revealing equilibria where prices fully reflect private information. Our findings mirror this result, because traders submit fully revealing priority fees. However, blockchains operate with periodic clearing, and the revealing equilibrium is delayed until the end of block time.

Trading in blockchains resembles trading in preopening sessions of traditional electronic exchanges, where markets clear only at the end of a short period. Unlike preopening sessions, where orders execute at a uniform clearing price that maximizes volume, blockchain orders are priced by priority fees, and volume is not maximized. Moreover, traders in preopening sessions observe partial information about the market, such as liquidity at the best prices or a virtual clearing price; see \cite{van2011transparency}.\footnote{\cite{biais1999price} propose several hypotheses regarding the informativeness of preopening prices; our empirical results in Appendix \ref{sec:instituionaldetails} show that information revelation is delayed in public memory pools, which supports their \textit{pure-noise hypothesis}, under which trading activity before clearing contains no new information. Similarly, \cite{medrano2001strategic} find that trading volumes and information revelation accelerate toward the end of preopening sessions.}

Our work is also related to the literature on the microstructure of DEXs.\footnote{See  \cite{hasbrouck2022need, angeris2021replicating2, capponi2023decentralized, milionis2022automated, cartea2023predictable, cartea2024decentralized} who show that DEXs cause losses to liquidity suppliers. \cite{drissi2025equilibrium} study DEXs as secondary trading venues. \cite{lehar2025decentralized} describe competition between DEXs and order books.    \cite{klein2023price} study the role of informed liquidity supply in price discovery. \cite{park2023conceptual} studies the different types of cost when trading in DEXs. \cite{hasbrouck2023economic} study concentrated liquidity DEXs. \cite{malinova2024learning} examine the potential of DEXs to organize trading for equity. \cite{capponi2024price} study the empirical influence of gas fees on the informativeness of trades. \cite{cartea2024strategic} propose AMM designs to mitigate losses of liquidity suppliers.} In contrast to this literature, we study DEXs as the interplay of a smart contract and the blockchain on which it operates. The economic tradeoffs for liquidity supply are similar to those in market making within limit order books and dealer markets, and insights from classical models apply (e.g., \cite{ho1983dynamics}, \cite{glosten1985bid}).  In particular, the trading function of a DEX can be viewed as a parameterized price schedule in electronic exchanges; see \cite{glosten1994electronic, biais2000competing}.  As a consequence, the conclusions of  our model apply if the blockchain runs other market infrastructures.  

\cite{he2025arbitrage} study the role of priority fees when arbitrageurs trade between a blockchain and a competing venue where prices are formed. In contrast, this paper studies theoretically whether the blockchain itself can serve as the primary venue for price formation.

Blockchain technology is profoundly reshaping the financial landscape (\cite{harvey2016cryptofinance, cong2019blockchain}), and  blockchain economics are a growing topic in the literature (\cite{biais2023advances} and \cite{petryk2025promises}). \cite{john2020proof} study the equilibrium security of blockchains. \cite{cong2021tokenomics} analyze the adoption of digital platforms. \cite{biais2023equilibrium} study the drivers of cryptocurrency prices and risks. \cite{cong2023scaling,harvey2025economic} study the economic implications of \emph{layer-2} blockchains. \cite{harvey2024international} explore how investors and regulators engage effectively with blockchains. Finally,  the applications of decentralized finance extend beyond DEXs. They include  asset tokenization (\cite{heines2021tokenization}), central bank digital currencies (\cite{fung2016central, auer2022central}), cross-border settlement (\cite{harvey2021defi, hub2023project, cardozo2024cross}).

The remainder of this paper proceeds as follows.
In Section~\ref{sec:Model}, we present notation,
assumptions, and structure of the model.
In Section~\ref{sec:solution stage three}, we study the
priority fee bidding strategies and trading
volumes of informed traders. In Section
\ref{sec:solution stage two}, we analyze the liquidity
supply. In Section~\ref{sec:solution stage one}, we study
incentives to acquire information at a cost. In
Section~\ref{sec:blocktime}, we study the economic
effects of block time. In Appendix  \ref{sec:instituionaldetails},
we describe the features of blockchains that underpin
our model. Finally,  Appendix \ref{apx:proofs} collects the proofs.

\section{General features of the model} \label{sec:Model}

In this section, we introduce the features of our model for the strategic interactions of market participants in blockchain-based markets. A decentralized exchange (DEX) supplies liquidity in a reference asset $X$ (e.g., dollar) and in a risky asset $Y.$\footnote{See Appendix \ref{apx:amms} and \cite{capponi2021adoption,cartea2024decentralized,lehar2025decentralized} for more details on the mechanics of DEXs.} There are three types of agent, a representative liquidity supplier (he), informed traders (she), and noise traders with price-sensitive demand. The entry to blockchain markets, liquidity supply, and the trading process are modeled as a game that proceeds as follows.
\begin{itemize}
    \item \emph{Stage one}: $M$ traders decide whether to pay a fixed information cost~$C$ to enter the blockchain and to observe private information. This choice depends on whether informed trading yields an expected future payoff higher than $C$.
    \item \emph{Stage two}: the liquidity supplier sets the liquidity reserves in the DEX by balancing expected losses to informed traders and expected fee revenue from noise traders. The level of reserves determines the cost of liquidity.
    \item \emph{Stage three}: Informed traders hold long-lived private and incomplete information about the future liquidation value of the asset.  Based on valuations, traders compete for queue priority in the next block, and they determine their priority fee bids and their trading volumes. In contrast, noise traders submit transactions with zero priority fees.\footnote{Builders are paid \emph{gas fees} by agents to include their transactions in a block. Gas fees consist of two components: the \textit{base fee} and the \textit{priority fee}. The base fee is mandatory for inclusion and is paid by all traders. We assume that noise traders pay the base fee only, which we normalize to zero without loss of generality, because the base fee does not affect ordering within the block. See Appendix \ref{apx:pgas} for further institutional details on the blockchain protocol.}    
\end{itemize}

After the three stages, a builder constructs the block by sorting transactions according to their priority fees. Specifically, the block executes the transactions of informed traders first, followed by those from noise traders who only pay the base fee.\footnote{In blockchains, informed traders do not profitably blend with noise traders to conceal their orders, as they would in \cite{kyle:1985}. Specifically, blending would require paying a zero priority fee. Submitting an arbitrarily small positive priority fee is a profitable deviation, because it secures a better queue position and execution price. In equilibrium, informed traders pay priority fees which reveal valuations; see Proposition~\ref{prop:eqpfwithvol}.}

After the liquidity supplier sets the DEX's liquidity in stage two, we assume for simplicity that he remains passive and does not compete with traders for queue priority in the subsequent block. Although the liquidity supplier does not compete directly, his strategy depends indirectly on the behavior of informed traders: he chooses the cost of liquidity in the DEX based on rational expectations about their best response in the next stage.

At the start of the PGA in stage three, each informed trader observes a private valuation, 
drawn from a common and known distribution. Next, they set priority fees and trading volumes by balancing expected profits against (i) the adverse price impact generated by traders with better positions in the block and (ii) slippage in the DEX.

A key assumption of the model is that all informed traders know whether the price innovation is positive or negative.
This determines the sign of all valuations and whether traders are buyers or sellers.\footnote{More precisely, the conditional information structure is as follows. At the beginning of the PGA, a public indicator reveals the direction of the price innovation, after which each informed trader observes a private and independent valuation regarding the magnitude of the innovation. As detailed below, positive valuations are drawn from a distribution with positive support, while negative valuations follow a symmetric distribution with negative support.} We motivate this assumption as follows. Informed traders submit their transactions
near the end of the blockchain slot and rely on private
information observed at that time.\footnote{Informed traders transact through private memory pools or wait until the end of block time in public memory pools, consistent with the empirical evidence 
reported in Appendix \ref{apx:pgas}.} Blockchains that host decentralized markets, such as Ethereum, operate with slots of twelve seconds or longer to build blocks securely.\footnote{Layer-two blockchains achieve shorter block times by relying on centralized sequencers, and thus do not qualify as  decentralized.} Thus, substantial price movements can occur during this period and cause most private valuations to take a definite sign. Our focus is on price discovery and liquidity in settings where the liquidation value of the asset is likely to change significantly in the DEX, so most informed traders are either buyers or sellers, and the PGA serves as the mechanism through which information is incorporated into prices.\footnote{In contrast, when the future liquidation value of the asset is close to its initial value, private valuations are more dispersed in direction. In such situations, if an informed buyer and an informed seller both submit transactions within the same block, each has an incentive for the other's trade to be executed first, because this improves their own execution price. Consequently, the expected value of queue priority becomes non-positive for both traders, leading to zero equilibrium bids and limited informational content in the auction outcome.}

We solve for a  perfect Bayesian equilibrium  of this game  by backward induction. At stage three (Section~\ref{sec:solution stage three}), given a known level of liquidity supply $L$ and a known  number $M$ of competing informed traders, the bidding strategies and trading volumes are determined. At stage two (Section~\ref{sec:solution stage two}), given a known  number $M$ of competing informed traders, the liquidity supply $L$ is determined based on the  supplier's rational expectations of future trading volumes and noise trading activity. At stage one (Section~\ref{sec:solution stage one}), the number $M$ of informed traders who pay the cost of information is determined. The sequence of events is described in more detail in the next sections. 

\section{Stage three: priority fees and trading volumes} \label{sec:solution stage three} 

This section studies the priority fee bidding strategies and trading volumes of informed traders who compete for queue priority in stage three. Traders take as given the liquidity supply in the DEX and the number $M$ of competitors. Assume that informed traders know the liquidation value $V$ is positive, so they wish to buy the asset. The case in which informed traders are sellers is symmetric and yields identical equilibrium priority fee bidding strategies, and  trading volumes of opposite sign. 

\subsection{Assumptions}

In stage three, $M$ risk-neutral informed traders compete for queue priority in the next block.  We assume that the DEX offers a linear price schedule, and implements a linear price update rule. Let $L$ denote the depth of liquidity in the DEX determined in stage two.\footnote{The depth $L$ is determined by the  by the reserves deposited in the DEX's liquidity pool, and by the shape of the \textit{bonding curve} implemented in the DEX's smart contract; see Appendix~\ref{apx:amms} for details.}   An order of size $Q$ is executed with slippage $Q/L$ per unit of the asset, and, after execution, the price updates linearly by $2\,Q/L$  in the direction of the trade. This assumption is a first-order approximation of the pricing mechanism in standard DEXs; see Appendix~\ref{apx:amms} for details.

Traders compete to avoid potentially worse execution prices resulting from the adverse price impact of preceding orders in the block. Thus, the supply $L$ influences competition outcomes because it determines execution prices and the magnitude of price impact. 
Consider two traders, $1$ and $2$, who sequentially buy quantities $Q_1>0$ and $Q_2>0$ of asset $Y$. We normalize the DEX's initial price at the start of the blockchain slot to zero, without loss of generality.  Thus, the execution price for trader $1$, assuming she has queue priority, is $Q_1/L$, after which the DEX price updates to $2\,Q_1/L$.  Consequently, the execution price for trader $2$, who trades second in the block, is $Q_2/L + 2\,Q_1/L$.\footnote{See Appendix~\ref{apx:amms} for more details on execution prices and price impact in DEXs.}

The PGA among traders is a multi-prize contest in which each competitor's payoff depends on the rank of her priority fee relative to others. As discussed in Appendix \ref{apx:pgas}, the PGA is \emph{blind} because traders either compete in the private memory pool or submit their transactions to the public memory pool immediately before block creation. Moreover, the PGA is \emph{all-pay} because traders' transactions are all executed and builders collect priority fees regardless of relative ranks within the block.

Before submitting their transactions, traders know whether they are all buyers or all sellers, and they observe private valuations $\{v_{1}, \ldots, v_{M}\}$ used to assess the asset's worth. Specifically, from the perspective of trader $i$, the expected liquidation value is 
\begin{equation}\label{eq:Vi}
\mathbb{E}_{i}\!\left[V\right] = v_i\,.
\end{equation}
Valuations are  independent and identically distributed (i.i.d.) with continuous cumulative distribution function $F$ on $[0,\overline v]$, admitting a continuous density $f$. In the perfect Bayesian equilibrium of the PGA, the $M$ risk-neutral traders determine their priority fees  \(\{\Phi_{1},\cdots,\Phi_{M}\}\) and their trading volumes \(\{Q_{1},\cdots,Q_{M}\}\).

\subsection{The priority gas auction}

This section derives the expected utility of competing traders in the PGA. The contest is symmetric, and we describe it from the perspective of trader $i \in \{1, \dots, M\}$. Trader $i$ competes with $M-1$ other traders in the PGA. From her perspective, the $M-1$ valuations of her competitors are i.i.d. draws from the same distribution; likewise, the $M-1$ trading volumes of her competitors are i.i.d. draws from the same distribution. We denote by $\Phi_{(j)}$, for $j \in \{1, \dots, M-1\}$, the $j$-th smallest bid among the $M-1$ competitors, so that $\Phi_{(j)} < \Phi_{(j+1)}$ and $\Phi_{(M-1)}$ is the largest competing priority fee. We denote by $v_{(j)}$ and $Q_{(j)}$ the valuation and trading volume, respectively, of the competitor submitting the priority fee ranked $j$-th.

If trader~$i$ wins priority in the block, i.e., if~$\Phi_i > \Phi_{(M-1)}$, her buy order is executed first in the next block at the price~$Q_i / L + \pi$ per unit of the asset, where $\pi$ is a proportional fee paid to the DEX; see~\eqref{eq:slippage}.\footnote{We assume that $\pi<\overline v<\infty$ to ensure that the probability of non-zero trading volumes is positive; see Section~\ref{sec:trading volumes}.} In this case, the trader's terminal wealth depends on the priority fee paid to the builder, the cash paid to the DEX to purchase~$Q_i$ units of the asset, the proportional DEX fee~$\pi$, the information cost~$C$ incurred in stage one, and the terminal value of her holdings:
\begin{equation}\label{eq:winwealth}
    W_{i,(M-1)}
    = \underbrace{-\Phi_i}_\text{priority fee}
    - \underbrace{Q_i \left( \frac{Q_i}{L} + \pi \right)}_{\text{cash paid to DEX}}
    + \underbrace{Q_i \,V}_{\text{terminal value of holdings}}
    - \underbrace{C}_{\text{information cost}}.
\end{equation}

If trader~$i$'s priority fee~$\Phi_i$ lies between the $j$-th and $(j+1)$-th smallest priority fees among her competitors, i.e.,
\[
    \Phi_{(j)} < \Phi_i < \Phi_{(j+1)}\,,
\]
then exactly~$M - 1 - j$ traders, with trading volumes~$\{Q_{(j+1)}, \dots, Q_{(M-1)}\}$, are executed in the block before trader~$i$'s buy order. The price impact in the DEX when trading a volume~$Q$ is~$2\,Q / L$. Hence, trader~$i$'s execution price per unit of the risky asset is
\[
    \underbrace{\frac{Q_i}{L}}_{\text{slippage}}
    + \underbrace{\frac{2}{L} \sum_{\ell = j + 1}^{M - 1} Q_{(\ell)}}_{\text{impact of traders with higher priority fees}}
    + \underbrace{\pi}_{\text{DEX fee}}\,.
\]
Accordingly, trader~$i$'s terminal wealth in this case is
\begin{align}\label{eq:losewealth}
    W_{i,(j)}
    &= \underbrace{-\Phi_i}_{\text{priority fee}}
       - \underbrace{Q_i \!\left( \frac{Q_i}{L}
       + \frac{2}{L} \Delta_{(j+1 : M-1)}
       + \pi \right)}_{\text{cash paid to DEX}}
       + \underbrace{Q_i V}_{\text{terminal value of holdings}}
       - \underbrace{C}_{\text{information cost}},
\end{align}
where, for convenience, we define
\[
    \Delta_{(l : L)} = \sum_{\ell = l}^{L} Q_{(\ell)}.
\]

Finally, if trader~$i$'s priority fee is lower than the lowest competing fee, i.e.,
\[
    \Phi_i < \Phi_{(1)}\,,
\]
then her terminal wealth is
\[
    W_{i,(0)}
    = -\Phi_{i}
      - Q_{i}\!\left(
        \frac{Q_{i}}{L}
        + \frac{2}{L}\,\Delta_{(1:M-1)}
        + \pi
      \right)
      + Q_{i} V
      - C\,.
\]

Using the cases above, the expected utility of trader $i$ can be written as
\begin{align*}
\mathbb{E}_{i}\!\left[W_{i}\right]
&= \sum_{j=0}^{M-1}\mathbb{E}_{i}\!\left[\mathbbm{1}_{\Phi_{(j)}<\Phi_{i}<\Phi_{(j+1)}}\,W_{i,(j)}\right] \nonumber \\
&= \sum_{j=0}^{M-1}\mathbb{E}_{i}\!\left[
    \mathbbm{1}_{\Phi_{(j)}<\Phi_{i}<\Phi_{(j+1)}}
    \Bigl(
        -\Phi_{i}
        -Q_{i}\!\left(\frac{Q_{i}}{L}
        + \frac{2}{L}\,\Delta_{(j+1:M-1)}
        + \pi\right)
        + Q_{i}V
        - C
    \Bigr)
\right]\!,
\end{align*}
with the conventions
\[
\Delta_{(l:L)}=0 \text{ whenever } l>L,\qquad \Phi_{(M)}=\infty,\qquad \Phi_{(0)}=-\infty.
\]

At stage three, trader~$i$'s valuation of the asset is $v_i$, and she knows the DEX supply $L$, the number of competing traders $M$, her trading volume $Q_i$, and her priority fee $\Phi_i$. She understands that her queue position depends on the rank of $\Phi_i$ relative to her competitors. Hence the expected wealth of trader~$i$ can be written as
\begin{equation}\label{eq:expwealth}
\mathbb{E}_{i}\!\big[W_{i}\big]
= -\Phi_{i}
+ Q_{i}\Big(v_{i} - \pi - \frac{Q_{i}}{L}\Big)
- C
- \frac{2\,Q_{i}}{L}
\sum_{j=0}^{M-1}
\mathbb{E}_{i}\!\left[
    \mathbbm{1}_{\{\Phi_{(j)}<\Phi_{i}<\Phi_{(j+1)}\}}
    \,\Delta_{(j+1:M-1)}
\right].
\end{equation}
The expected wealth~\eqref{eq:expwealth} consists of (i) the expected profit~$Q_i\, v_i$ and (ii) several trading costs. These costs include the priority fee paid to builders, the DEX fee, the information cost, the expected adverse price impact generated by trader~$i$'s competitors, and the execution price paid to the DEX.

When submitting a transaction, trader $i$ determines both the priority fee and the volume, thus she solves the problem\footnote{ The order of the two suprema can be interchanged.}
\begin{equation}\label{eq:optimisation}
\sup_{Q_i}\,\sup_{\Phi_i}\mathbb{E}_i\left[W_{i}\right]\,.
\end{equation}

Intuitively, traders buy and sell volumes proportional to their private valuations of the asset. In the optimization problem \eqref{eq:optimisation}, the trading volume $Q_i$ for trader $i$ depends on $v_i$. Similarly, the priority fee $\Phi_i$  depends on the trading volume $Q_i$. Thus, in equilibrium, both the volume $Q_i$ and the priority fee $\Phi_i$ are functions of the private valuation $v_i\,.$\footnote{Specifically, note that $\partial_{Q_{i}}\mathbb{E}_{i}\left[W_{i}\right]$ is a function $v_i,$ i.e., $\partial_{Q_{i}v_{i}}\mathbb{E}_{i}\left[W_{i}\right]\ne 0$. Thus, the optimal trading volume $\argmax_{Q_{i}}\mathbb{E}_{i}\left[W_{i}\right]$ is a function of the valuation $v_i$. Similarly, $\partial_{\Phi_{i}v_{i}}\mathbb{E}_{i}\left[W_{i}\right]\ne 0$, so the priority fee also depends on the private valuation.}
Section~\ref{sec:priority fees} derives the equilibrium priority fees for a fixed distribution of trading volumes, and Section~\ref{sec:trading volumes} then derives the equilibrium trading volumes given these priority fees.

\subsection{Priority fees}
\label{sec:priority fees}

Here, we solve for the equilibrium priority fee under an arbitrary distribution of trading volumes. Let $Q_j$ denote the random variable representing the trading volume of trader $j\ne i$, drawn from the interval $[0, \overline{Q}]$ according to a density function $g$ with finite first and second moments, which may include an atom $p = G\!\left(\{0\}\right)$ at zero volume. We denote by $G$ the cumulative distribution function of volumes.
At zero volume, traders submit the minimum zero priority fee. Equivalently, they do not participate in the PGA, and their expected wealth equals $-C$.\footnote{In particular, no tie-breaking rule is required for zero volumes. Note also that a zero trading volume generates no adverse price impact on subsequent transactions within the block.}  The equilibrium distribution of trading volumes is derived in Section~\ref{sec:trading volumes}.

When solving for the equilibrium at stage three, informed traders take the depth~$L$ as fixed and independent of the priority fee~$\Phi_i$. Although the priority fee does not directly affect liquidity supply, it exerts an indirect influence. As shown in Section~\ref{sec:solution stage two}, the liquidity supplier anticipates the priority fees and trading volumes of informed traders when determining the depth of the DEX's pool.

To better characterize the economic object of competition in the PGA, we rewrite trader~$i$'s expected wealth in~\eqref{eq:expwealth} as
\begin{equation}\label{eq:expwealth2}
\mathbb{E}_{i}\!\left[W_{i}\right]
= \mathbb{E}_{i}\!\left[W_{i,(0)}\right]
+ \frac{2\,Q_{i}}{L}
\sum_{j=0}^{M-1}
\mathbb{E}_{i}\!\left[
    \mathbbm{1}_{\{\Phi_{(j)} < \Phi_{i} < \Phi_{(j+1)}\}}
    \,\Delta_{(1:j)}
\right],
\end{equation}
which has the following economic interpretation. The term $\mathbb{E}_{i}\!\left[W_{i,(0)}\right]$ represents trader~$i$'s expected wealth if she is executed last in the block. By submitting a positive priority fee~$\Phi_i$, she obtains, with positive probability, an additional surplus. This surplus,
\begin{equation}\label{eq:prizevalue}
\frac{2\,Q_{i}}{L}
\sum_{j=0}^{M-1}
\mathbb{E}_{i}\!\left[
    \mathbbm{1}_{\{\Phi_{(j)} < \Phi_{i} < \Phi_{(j+1)}\}}
    \,\Delta_{(1:j)}
\right],
\end{equation}
represents the expected price impact that trader~$i$ avoids by improving her queue position. The magnitude of this surplus increases with the rank of~$\Phi_i$ among competitors and with the quantity demanded by the trader.

Trader $i$ expects her competitors to employ a continuous and differentiable priority fee bidding strategy $\Phi(\cdot)$ which is increasing in the trading volume $Q$. Therefore, the probability of submitting the $j+1$-th largest priority fee is
$$
\mathbb{P}_{i}\left[\Phi_{\left(j\right)}<\Phi_{i}<\Phi_{\left(j+1\right)}\right]=\mathbb{P}_{i}\left[Q_{\left(j\right)}<\Phi^{-1}\left(\Phi_{i}\right)<Q_{\left(j+1\right)}\right]\,.$$
To determine her optimal priority fee, trader $i$ solves the optimization problem
\begin{equation}
\sup_{\Phi_{i}}\mathbb{E}_{i}\left[W_{i}\right]\,,
\end{equation}
which is equivalent to solving
\begin{equation}\label{eq:optimisation1}
\sup_{\Phi_{i}}\left\{ -\Phi_{i}+\frac{2\,Q_{i}}{L}\,\sum_{j=0}^{M-1}\mathbb{E}_{i}\left[\mathbbm{1}_{\Phi_{\left(j\right)}<\Phi_{i}<\Phi_{\left(j+1\right)}}\Delta_{\left(1:j\right)}\right]\right\} \,.
\end{equation}
That is, trader~$i$ chooses her priority fee to maximize the sum of (i) the loss from paying the priority fee to the builder and (ii) the gain from the surplus. Trader~$i$ therefore faces a tradeoff: lowering the priority fee increases expected wealth but decreases the probability of attaining a higher position in the queue.

To further analyze the optimization problem, the following lemma is useful, because it simplifies trader~$i$'s estimation of the expected surplus from the PGA.
\begin{lem}\label{lem:simplifiedExpectedImpact}
Assume $M\ge 2$. The surplus \eqref{eq:prizevalue} is
\begin{align}
\frac{2\,Q_i}{L}\,\sum_{j=0}^{M-1}\mathbb{E}_{i}\left[ \mathbbm{1}_{\{\Phi_{(j)}<\Phi_{i}<\Phi_{(j+1)}\}}\,\Delta_{(1:j)}\right]   = \frac{2\,Q_i}{L}\,\left(M-1\right)\int_{0}^{\Phi^{-1}\left(\Phi_{i}\right)}x\,dG\left(x\right)\,.
\end{align}
\end{lem}

Lemma~\ref{lem:simplifiedExpectedImpact} shows that the surplus is obtained by averaging the price impact of transactions with volumes lower than trader~$i$'s inverse bid. This result follows from the  structure of the PGA. First, it arises from our assumption of additive and linear price impact in the DEX. In practice, nonlinearities exist in the impact of consecutive trades; see~\cite{cartea2023predictable}. However, we model competition within a block whose duration is short enough for our linear approximation \eqref{eq:impact} to be accurate. Second, the result of Lemma~\ref{lem:simplifiedExpectedImpact} also relies on the strict ordering of transactions by priority fee within the block. In practice, while builders have reputational incentives to maintain strict ordering, they are not required to do so by the blockchain protocol. Such strategic considerations are beyond the scope of this work.

In a Bayesian-Nash equilibrium, trader $i$ finds it optimal to adopt the same differentiable and increasing strategy $\Phi$, which pins down the equilibrium strategy. The solution is characterized in the following result.
\begin{proposition}
\label{prop:pf}
The equilibrium priority fee is
\begin{equation}\label{eq:priority fee eq model 1}
\Phi\left(Q_{i}\right)=\frac{2}{L}\left(M-1\right)\int_{0}^{Q_{i}}x^{2}\,dG\left(x\right)\,.
\end{equation}
The priority fee is increasing in the trading volume $Q_i$, increasing in the number of informed traders $M$, and decreasing in the liquidity depth $L.$ Finally, trader $i$'s objective \eqref{eq:optimisation1} is
\begin{equation}\label{eq:eqobjpf}
\frac{2}{L}\,\left(M-1\right)\int_{0}^{Q_{i}}x\,\left(Q_{i}-x\right)\,dG\left(x\right)\,.
\end{equation}
\end{proposition}

To set her priority fee, trader~$i$ uses the reservation priority fee
\[
\frac{2\,Q_{i}}{L}\,(M-1)\int_{0}^{Q_{i}} x\,dG(x)
\]
as a baseline when computing her optimal fee. This reservation priority fee is the level that makes trader~$i$ indifferent between participating in the PGA and abstaining, that is, it sets her objective~\eqref{eq:eqobjpf} equal to zero. Trader~$i$ then applies a discount to this baseline to capture a surplus. In particular, as trader~$i$'s trading volume~$Q_i$ increases, the likelihood that her reservation fee exceeds those of other traders also rises. Consequently, trader~$i$ applies a larger discount as the value of her private volume increases.

For a fixed arbitrary distribution of trading volumes, equilibrium priority fees increase with the number of competitors. Specifically, as the number of informed traders rises, the expected adverse price impact becomes larger, incentivizing each trader to raise her priority fee. Conversely, equilibrium priority fees decrease with liquidity supply. Specifically, when $L$ is large, price impact is low and the cost of losing queue priority becomes less significant.

As we show below, however, these relationships do not necessarily hold (i) when informed traders choose their trading volumes strategically as a function of anticipated priority fees, and (ii) when liquidity suppliers set their supply strategically as a function of anticipated trading volumes.

\subsection{Trading volumes and participation cutoff}
\label{sec:trading volumes}

The section above derived the equilibrium priority fees of informed traders for an arbitrary distribution~$g$ of trading volumes with support~$\left[0, \overline{Q}\right]$. We now endogenize this distribution by determining the volumes that maximize the informed traders' objective. Substituting the optimal priority fee~$\Phi_i$ from~\eqref{eq:priority fee eq model 1} into the objective function~\eqref{eq:optimisation}, we express trader~$i$'s optimization problem as
\begin{equation}
\label{eq:optimisationvolumes}
\sup_{Q_{i}}\Bigg\{
-\frac{2}{L}\,(M-1)\left(\int_{0}^{Q_{i}}x^{2}\,dG(x)
+ Q_{i}\int_{Q_{i}}^{\overline{Q}}x\,dG(x)\right)
+ Q_{i}\left(v_{i} - \pi - \frac{Q_{i}}{L}\right)
\Bigg\}\,.
\end{equation}

The programme \eqref{eq:optimisationvolumes} is interpreted as follows. In equilibrium, all traders employ the same priority fee bidding strategy, allowing them to estimate, on average, the expected costs of the PGA in the form of expected adverse price impact and priority fees. Additional trading costs incurred by trader $i$  include (i) the slippage $-Q_i^2/L$, (ii) the DEX's   fee $-\pi\,Q_i$, and (iii) the cost of information $-C$. Both trading costs and infrastructure costs increase with trading volume, creating an incentive for trader $i$ to reduce her trading volume $Q_i$. However, trader $i$'s expected trading profits are $Q_i\,v_i$. This provides an incentive for trader $i$ to increase her holdings in the asset.

We look for equilibria in which trading volumes are a weakly monotone function of private valuations and we write $Q_i = Q(v_i)$ for $v_i \in [0, \overline v].$ In particular, the minimum and maximum trading volumes satisfy
\[
Q(0) = 0
\qquad \text{and} \qquad
\overline{Q} = Q(\overline{v})\,.
\]
We later verify that the monotonicity property holds in equilibrium.  The distribution of trading volumes, expressed as a function of $F$, is given by
\begin{equation}\label{eq:defg}
G(x) = F\!\big(Q^{-1}(x)\big)\,,
\end{equation}
where $Q^{-1}(\cdot)$ denotes the generalized inverse of the function $Q(\cdot)$. 

The adverse price impact of competing traders depends on trader~$i$'s private valuation of the asset. Lower valuations correspond to lower volumes and to worse queue positions within the block, and consequently, to higher adverse price impact from traders with better positions. For sufficiently low valuations, the adverse price impact associated with high-valuation traders may discourage trader~$i$ from submitting a transaction if doing so results in a negative objective~\eqref{eq:optimisationvolumes}. In what follows, we show that there indeed exists a participation cutoff~$\underline{v}_M$ for private valuations below which trader~$i$ optimally chooses not to submit a transaction. 

Using \eqref{eq:defg}, the first-order condition (FOC) derived from the optimization problem \eqref{eq:optimisationvolumes} yields the following implicit integral equation for trader $i$'s optimal trading volume
\begin{equation}\label{eq:integralEqVolume2}
\left(v_{i}-\pi\right)L - 2\,Q(v_i) - 2\,(M-1)\int_{v_i}^{\overline{v}} Q(x)\,dF(x) = 0\,.
\end{equation}

Next, Lemma \ref{lem:cutoff} characterizes the participation cutoff, and Proposition \ref{prop:volume} derives the equilibrium trading volumes and their distribution.

\begin{lem}\label{lem:cutoff}
Fix $M\ge 2$. There exists a unique $\underline v_M\in(0,\overline v)$ which satisfies
\begin{equation}\label{eq:cutoff}
\underline v_M - \pi - \int_{\underline v_M}^{\overline v}\bigl(e^{(1-F(u))(M-1)} - 1\bigr)\,du = 0\,.
\end{equation}
Moreover, $\underline v_M$ is strictly increasing in $M$ and $\lim_{M\rightarrow\infty}\underline v_M = \overline v.$
\end{lem}

\begin{proposition}\label{prop:volume}
Fix $M\ge 2$. If $v_i<\underline v_M$, where $\underline v_M$ is defined in \eqref{eq:cutoff}, then trader $i$'s optimal trading volume is $Q(v_i)=0$. If $v_i\ge\underline v_M$, then the optimal trading volume is proportional to the liquidity supply $L$,
\begin{equation}\label{eq:eqvolume}
Q(v_i)=L\,\tilde{Q}(v_i)\,,
\end{equation}
where
\begin{equation}\label{eq:eqvolumetilde}
\tilde{Q}(v_i)
= \frac{1}{2}\Bigg(\left(\overline v - \pi\right)e^{-(M-1)\left(1-F(v_i)\right)}
- \int_{v_i}^{\overline v} e^{-(M-1)\left(F(u)-F(v_i)\right)}\,du\Bigg).
\end{equation}
Moreover,
\[
Q(\underline v_M) = 0 \qquad\text{and}\qquad
\overline{Q} = Q(\overline v) = L\frac{\overline v - \pi}{2}.
\]
The trading volume $Q(v_i)$ is increasing in the private valuation $v_i$, decreasing in the number of informed traders $M$ with limit $\lim_{M\to\infty}Q(v_i)=0$ for every $v_i<\overline v$.
\end{proposition}

Lemma~\ref{lem:cutoff} shows that there is a participation cutoff $\underline v_M$ for private valuations: informed traders with valuations below $\underline v_M$ find it unprofitable to trade and abstain from submitting a transaction. The cutoff is strictly increasing in the number of competitors $M$ and converges to the upper bound $\overline v$ as $M\to\infty$. 

When a trader's valuation exceeds the participation cutoff~$\underline v_M$, Proposition~\ref{prop:volume} shows that her trading volume is proportional to the liquidity supply~$L$, with proportionality factor~$\tilde{Q}(v_i)$. Thus, deeper liquidity reserves allow traders to extract greater profits from private information. In contrast, trading volumes decrease with the number of traders in anticipation of higher expected PGA costs.

\begin{figure}[!htb]
\centerline{
\includegraphics[]{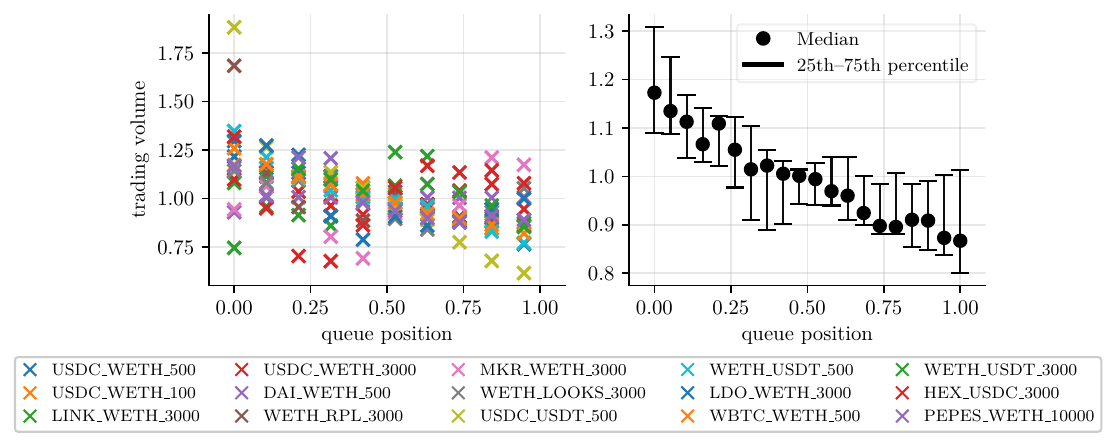}
}
  \caption{Left panel: Average absolute trading volume as a function of queue position in the block for transactions in multiple Uniswap v3 pools; $0$ corresponds to first position and $1$ to last position. The transactions  are between $1$ January $2023$ and $31$ December $2023$ in $15$ different Uniswap v3 pools with multiple transactions in at least $10$ different blocks. For each pool, the trading volumes are normalized by the standard deviation of trading volumes. Right panel: average and inter-quartile trading volumes across the $15$ pools.} \label{fig:trading volumes}
\end{figure}

Moreover,  the factor~$\tilde{Q}(v_i)$ increases with the private valuation~$v_i$. This occurs for two reasons. One, a higher valuation strengthens the incentive to trade, and two, it increases the expected surplus from the PGA. Thus, higher private valuations correspond to larger trading volumes, and as shown in Proposition \ref{prop:pf}, they are also associated with better queue positions. To test this prediction, Figure \ref{fig:trading volumes} shows the average trading volume across multiple DEX pools on Ethereum as a function of their queue position within the block. The figure indicates that the  volume of transactions decreases with queue position, and lends empirical support to the findings of our model.
\
\

Proposition~\ref{prop:volume} also shows that the participation cutoff~$\underline v_M$ for profitable informed trading increases with~$M$. Therefore, as the number of competitors grows, increasingly higher valuations are required to participate. Because only high-valuation traders ultimately submit transactions in blockchains, higher trading volumes, associated with higher valuations, should be more frequently observed on blockchains. This prediction is consistent with empirical evidence showing that per-transaction trading volumes in blockchain markets are significantly higher than those in traditional electronic limit order books. For instance, between 1~July~2021 and 31~December~2023, the average size of liquidity-taking trades in the most active DEX on Ethereum was approximately~\$70{,}000, compared with about~\$1{,}200 for the same pair of assets on Binance; see  \cite{cartea2025decentralised}.

\paragraph{Competition and price efficiency.} 
Proposition~\ref{prop:volume} shows that the equilibrium distribution of trading volumes is a transformation of the truncated distribution of valuations over~$[\underline v_M, \overline v]$:
\begin{equation}\label{eq:distGeq}
G\left(\{0\}\right) = F(\underline v_M)
\qquad \text{and} \qquad
G\big(Q(v)\big) = F(v)
\quad \text{for } v \in [\underline v_M, \overline v]\,.
\end{equation}
For a given number~$M$ of competing traders, the expected number of traders with sufficiently high valuations for the asset to submit non-zero trading volumes is~$M\big(1 - F(\underline v_M)\big)$. We refer to these traders as \emph{active traders}. 

In what follows, we assume that the probability density function of valuations $f$ does not exceed the uniform density in a (left) neighborhood of $\overline v$.\footnote{The upper bound $\overline v$ on valuations can be taken arbitrarily large to accommodate extreme valuations and ensure right-tail decay, so the condition holds for most standard distributions on compact support.}
The next result characterizes the behavior of price and volume in blockchain markets as competition intensifies.

\begin{proposition}\label{prop:activetraders}
The expected aggregate trading volume of active informed traders admits the closed form
\begin{equation}\label{eq:aggvolclosedform}
L\,M \int_{\underline v_M}^{\overline v}\tilde{Q}(x)\,dF(x) \;=\; \frac{L\,M\,(\underline v_M - \pi)}{2\,(M-1)}\,.
\end{equation}
The aggregate trading volume admits the limit $L\,\dfrac{\overline v-\pi}{2} =L\, \tilde Q( \overline v)$  as competition intensifies ($M\to\infty$).
Moreover, the expected price at the end of the block admits the closed form 
\begin{equation}\label{eq:priceclosedform}
\dfrac{M\,(\underline v_M-\pi)}{M-1}
\end{equation}
and converges to $\overline v - \pi$. Both limits are approached from below.
\end{proposition}

Proposition~\ref{prop:activetraders} shows that, as competition intensifies, a growing mass of active traders with increasingly high valuations participate in the PGA. For instance, when valuations are uniformly distributed, the participation cutoff admits the form $\underline v_M=\overline v-(\overline v-\pi)\log M/(M-1)$, so the range $[\underline v_M,\overline v]$ of active traders' valuations shrinks at rate $\log M/M$. Yet the expected number of active traders $M(1-F(\underline v_M))=M\log M/(M-1)$, the aggregate trading volume in~\eqref{eq:aggvolclosedform}, and the expected price at the end of the block in~\eqref{eq:priceclosedform} all  increase strictly with $M$.

In the limit when $M\to\infty$, aggregate volume converges to the volume $L\,\tilde{Q}(\overline v)$ associated with an informed trader holding the maximal valuation $\overline v$. Moreover, the expected price at the end of the block converges to $\overline v - \pi$, which corresponds to the upper bound of the valuation support net of fees.

The economic mechanism behind our result is endogenous selection. As $M$ increases, participation requires progressively more aggressive bidding to remain competitive. Traders anticipate both higher expected priority fees and the adverse price impact generated by participants with larger valuations. This raises the cutoff $\underline v_M$ and excludes traders with moderate valuations from the market. In equilibrium, only the most aggressive traders remain active. Their aggregate trading volume therefore reflects valuations arbitrarily close to $\overline v$, pushing the execution price toward its maximal feasible level net of fees.

These results have direct implications for price efficiency on a blockchain that serves as the venue for price formation. From the perspective of an uninformed observer, only non-zero priority fees and executed volumes reveal private information. As competition increases, the fraction of the valuation distribution that generates transactions shrinks. Price formation is thus driven by an increasingly thin upper tail of the distribution. Thus, the transaction price becomes systematically biased upward (downward if traders are selling).

As we discuss below, an increase in block time, that is, the time required for a block to be created and markets to clear, may raise the upper bound of possible valuations $\overline v$ and amplify the upward price bias. These effects play a role in shaping the equilibrium supply of liquidity, which we characterize in Section~\ref{sec:solution stage two}.

\subsection{Equilibrium priority fees}

For a fixed arbitrary distribution of trading volumes, Proposition \ref{prop:pf} showed that higher competition in the PGA increases priority fees and that deeper liquidity decreases them. However, trading volumes are part of the traders' equilibrium strategy. The next result derives the equilibrium priority fee accounting for strategic trading volumes.

\begin{proposition}\label{prop:eqpfwithvol}
The equilibrium priority fee is zero if $v_i< \underline v_M$. For $v_i\ge\underline v_M$, the equilibrium priority fee accounting for strategically adjusted trading volumes,  is
\begin{equation}\label{eq:eqpfwithvol}
\Phi\left(v_{i}\right)=2\,L\left(M-1\right)\int_{\underline{v}_M}^{v_{i}}\tilde{Q}\left(x\right)^{2}\,dF\left(x\right)\,.
\end{equation}
The equilibrium priority fee is  increasing in the  depth of liquidity $L,$ and, for every $v_i<\overline v$, it admits the limit
$$
\lim_{M\rightarrow\infty} \Phi(v_i) = 0\,.
$$
\end{proposition}

Proposition \ref{prop:eqpfwithvol} shows that the equilibrium priority fees $\Phi(Q(v))$, accounting for strategically adjusted  volumes, converge monotonically to zero with the number of informed traders in the memory pool. The intuition is that increased competition raises anticipated PGA costs and expected adverse price impact, which induces a down scaling  in equilibrium trading volumes. As $M \to \infty$, both trading volumes and priority fees converge to zero.

Similar arguments to those above show that the equilibrium priority fees and trading volumes derived earlier also apply when traders wish to sell the asset. As a consequence, traders are symmetrically aggressive in equilibrium, i.e., they buy and sell the same volumes for valuations of equal absolute value and opposite signs.

\section{Stage two: liquidity supply}\label{sec:solution stage two}

\subsection{Assumptions}

The level of reserves deposited in the DEX determines the execution costs and the price impact of liquidity taking trades.  At stage two, a risk-neutral representative liquidity supplier sets the DEX reserves by balancing losses to informed traders with fee revenue from noise traders.\footnote{Unlike traditional venues such as limit order books or dealer markets, the permissionless nature of blockchains facilitates entry to liquidity provision, particularly by less sophisticated participants. As a result, strategic liquidity suppliers cannot freely set the price of liquidity because of \emph{noise liquidity suppliers}. This feature of blockchains may explain the losses observed in practice; see \cite{cartea2024decentralized}. } Liquidity supply is competitive and is determined by a zero-profit condition: expected fee revenue from noise traders equals expected losses to informed traders.\footnote{This views the representative liquidity supplier as a continuum of competitive, atomistic, price-taking liquidity suppliers operating under free entry.}

For simplicity, we assume that noise traders transact a net volume that sums to zero in expectation, but an absolute expected volume which is price-sensitive. Specifically, the liquidity demanded by buyers is decreasing in the price $1/L$ and its expected volume is
\begin{equation*}
\frac{N}{2}\times\frac{1}{1+\theta/L}\,,
\end{equation*}
and the expected liquidity demanded by sellers is symmetric.

Demand on each side is therefore characterized by two parameters: $N/2$, the aggregate positive (resp.~negative) liquidity needs over the blockchain slot, and $\theta$, which governs the sensitivity of realized demand to liquidity depth $L$. When $L \ll \theta$, demand responds approximately linearly to liquidity, with slope proportional to $1/\theta$. As $L$ increases, marginal gains from additional liquidity diminish. Smaller values of $\theta$ imply that most of the available demand is captured at relatively modest depth, whereas larger values imply that demand is more sensitive and requires substantial liquidity to materialize. Our specification is bounded, so realized demand does not exceed the available mass $N$. Liquidity demand follows, in spirit, the reduced-form models of demand in \cite{garman1976market, ho1981optimal, hendershott2014price}. The liquidity supplier's expected fee revenue is therefore
\begin{equation}
\pi N \frac{L}{L+\theta}.
\end{equation}

The liquidity supplier does not hold the same information about the future liquidation value $V$ of the asset as that of informed traders. At stage two, he assumes that with probability $1/2,$ the liquidation value $V$ of the asset is positive, in which case the informed traders buy the asset and their private valuations are drawn independently from the interval $[0, \overline{v}]$ according to the continuous and differentiable density $f$. Similarly,  with probability $1/2,$ the liquidation value of the asset is negative, in which case the informed traders sell the asset and their valuations are drawn independently from the interval $[-\overline{v}, 0]$, according to the symmetric density $f(-x)$.

Recall that only traders who value the asset above the cutoff \eqref{eq:cutoff} submit transactions with volumes \eqref{eq:eqvolume}. We denote the total trading volume of informed traders by
\begin{equation}
\Delta_M = \sum_{k=1}^M Q(v_k)\,.
\end{equation}
From the perspective of the supplier, $\Delta_M$ is a random variable, whose expected value is zero, and which is assumed to be independent of the noise liquidity demand.

The liquidity supplier balances fee revenue with expected losses to informed traders. Next, we derive the expected adverse selection cost. Let $X_0$ and $Y_0$ denote the initial cash and risky asset reserves deposited by the liquidity supplier, so his initial wealth is $X_0 + Y_0\,V_0 = X_0$. Under the DEX's linear price schedule, the aggregate informed flow $\Delta_M$ executes at price $\Delta_M/L$. The liquidity supplier's cash therefore changes by $\Delta_M\,(\Delta_M/L) = \Delta_M^2/L$.\footnote{The change is similar whether informed traders buy or sell.} The liquidity supplier marks his position at the end of the block, after trading unfolds, at the prevalent DEX price $2\Delta_M/L$.\footnote{Each infinitesimal liquidity supplier marks wealth at the DEX's price because the DEX is the only venue to unwind inventory. Thus, a cash-out trade executes at the marginal price.}. Thus, his wealth becomes
\[
X_0 \;+\; \frac{\Delta_M^2}{L} \;+\; (Y_0-\Delta_M)\,\frac{2\Delta_M}{L},
\]
and the change in wealth attributable to informed trading is
\begin{equation}\label{eq:change wealth informed trading}
\frac{\Delta_M^2}{L} \;+\; (Y_0-\Delta_M)\,\frac{2\Delta_M}{L}
\;=\; -\,\frac{\Delta_M^2}{L} \;+\; \frac{2\,Y_0\,\Delta_M}{L}.
\end{equation}

The first term in \eqref{eq:change wealth informed trading} is the adverse selection cost. It is non-positive because the liquidity supplier holds more reserves of the asset whose price has decreased, or less of the asset whose price has increased.  The second term in \eqref{eq:change wealth informed trading} is the change in value of the liquidity supplier's initial inventory. Under the prior $\mathbb E[\Delta_M]=0$, the term has zero expectation. Hence the expected change in the  wealth attributable to informed trading at stage three is
\begin{equation}\label{eq:loss lp 0}
- \frac1L \, \mathbb E\left[\Delta_M^2\right] = -L\,M\,S_M\,,
\end{equation}
where we define
\begin{equation}\label{eq:defSM}
S_M = \int_{\underline v_M}^{\overline v} \tilde{Q}(u)^2\,dF(u)\;+\;\frac{(\underline v_M-\pi)^2}{4\,(M-1)}\,,
\end{equation}
and the volume function $\tilde Q$ is defined in \eqref{eq:eqvolume}.\footnote{The first term in \eqref{eq:defSM} is the expected squared volume of an active trader. The second term reflects the positive correlation between any two traders' volumes. Equations \eqref{eq:loss lp 0} and \eqref{eq:defSM} are derived in the proof of Proposition~\ref{prop:liquidity supply}.} 

\subsection{Equilibrium liquidity supply}

The aggregate equilibrium liquidity supply $L^\star$ satisfies a zero-profit condition: expected fee revenue from noise traders equals expected losses to informed traders.
\begin{equation}\label{eq:optimisation problem LP}
\pi\,N\,\frac{L^\star}{L^\star+\theta} = L^\star\,M\,S_M\,.
\end{equation}
The next proposition derives the equilibrium DEX liquidity supply.

\begin{proposition}\label{prop:liquidity supply}
In equilibrium, the supply of liquidity is
\begin{equation}\label{eq:liq supp eq}
L^\star(M)=\frac{\pi\,N}{M\,S_{M}}-\theta\,,
\end{equation}
where $S_M$ is defined in \eqref{eq:defSM}.  Liquidity supply increases in the profitability $\pi\,N$ of noise trading flow, and admits the limit
\begin{equation}\label{eq:limit kappa}
\lim_{M\rightarrow\infty} L^\star(M) = \frac{8\,\pi\,N}{3\,(\overline{v}-\pi)^2}-\theta\,,
\end{equation}
which is approached from above. The condition for markets to remain open is
\begin{equation}\label{eq:profitability condition}
 M\,S_{M}\le\frac{\pi\,N}{\theta}\,.
\end{equation}
\end{proposition}

Trading volumes in stage three are strategically set by informed traders to be proportional to the liquidity depth $L$. Thus, greater liquidity provision in stage two amplifies, on average, the dollar value of informed trading profits and, consequently, the expected losses \eqref{eq:loss lp 0} of the liquidity supplier. The liquidity supplier anticipates that the dispersion $S_M$ of aggregate trading volumes determines these losses and decreases supply accordingly. Conversely, he anticipates that noise demand is profitable and increases supply in response.

We now discuss how liquidity supply depends on the degree of competition in the PGA. As the number of competitors $M$ increases, the aggregate trading volume~\eqref{eq:aggvolclosedform} and the expected price~\eqref{eq:priceclosedform} at the end of the block both rise toward their limits (Proposition~\ref{prop:activetraders}). As a consequence, the expected adverse selection cost per unit of supplied liquidity also rises. The liquidity supplier anticipates this and decreases DEX liquidity. In the limit when $M\to\infty$, the price is most biased, liquidity is smallest, and only a small portion of private information is revealed. When valuations are uniformly distributed, liquidity supply admits the closed form
\begin{equation}\label{eq:liquidity uniform}
L^\star(M) \;=\; \frac{8\pi N(M-1)^3}{M(\overline v-\pi)^2\Bigl[(M-1)(3M-5)-2(2M-3)\log M+2(\log M)^2\Bigr]}-\theta,
\end{equation}
which is strictly decreasing in $M$.

Our results stand in sharp contrast to the effect of competition among informed traders in traditional markets with continuous trading and multiple clearing rounds. In such markets, as the number of informed traders tends to infinity, prices become fully revealing in early rounds and adverse selection premia vanish later; see \cite{holden1992long}. Market depth increases as private information is incorporated through successive rounds of trading.

The mechanism behind our result is fundamentally different. In blockchains, trades are executed in a single discrete clearing round and priced according to priority fees. Traders pay both the cost of market impact and the priority bid required to secure queue position. Because clearing occurs in one round, the liquidity supplier must absorb the full adverse selection risk from aggregate trading volume in a single decision, rather than adjusting liquidity across successive rounds.

\subsection{Market viability}

Under the zero-profit condition \eqref{eq:optimisation problem LP}, the equilibrium liquidity supply $L^\star$ is nonnegative only if expected fee revenue from noise traders can cover expected losses to informed traders. This requires the condition \eqref{eq:profitability condition} for the viability of blockchain markets. Condition \eqref{eq:profitability condition} is standard, it states that if the fee revenue from liquidity demand is less than the expected losses to informed traders,  liquidity provision is unsustainable. The degree to which this condition must hold depends on the price-sensitivity parameter $\theta$. All else being equal, as demand becomes more sensitive to the price of liquidity, the viability of liquidity provision requires either greater liquidity demand $N$, higher fee rates $\pi$, or reduced variance in informed trading volumes. This condition mirrors that in \cite{glosten1985bid} where markets shut down due to liquidity freeze if the valuation of informed traders is too precise relative to the price-sensitivity of liquidity demand.

When the variance of trading volumes is maximal, i.e., when
\begin{equation}\label{eq:maxvol}
S_M = \frac{\pi\,N}{M\,\theta}\,,
\end{equation}
then the liquidity depth is $L^{\star}=0$.

Our analysis in this section treats the number $M$ of informed traders as exogenous. In the next section, we endogenize entry into blockchains. The equilibrium number of competing traders in the PGA is constrained by the cost of acquiring information and by the limited profitability induced by defensive liquidity supply.

\section{Stage one: information acquisition}\label{sec:solution stage one}

Our results in Section~\ref{sec:solution stage two} show that, upon observing their private valuations, traders with low valuations submit zero trading volumes, while traders with high valuations participate in the PGA. At stage one, however, traders must decide whether to incur the information cost~$C$ to acquire private information before observing their valuations. These costs include expenditures on proprietary blockchain data, centralized exchange data, searcher contract deployment, and blockchain monitoring tools.

Assume that the total number of traders eligible to acquire information is arbitrarily large, and let  $M$ denote the subset that chooses to do so. In equilibrium, entry into informed trading is determined by the comparison between the expected profitability of informed trading and the information cost $C$. At stage one, the equilibrium number of informed traders
$M$ is the largest integer value such that the marginal trader's expected profit is at least the information cost. We characterize this equilibrium number in the following result.
\begin{proposition}\label{prop:eq nb traders}
The equilibrium number of informed traders is the largest integer $M$ such that
\begin{equation}\label{eq:eqconditionM}
C\;\le\;H(M)\,,\qquad{H(M)\;:=\;L^{\star}(M)\!\int_{\underline v_M}^{\overline v}\!\tilde Q(v)^2\bigl[1-2(M-1)(1-F(v))\bigr]\,dF(v),}
\end{equation}
where $\underline v_M$ and $\tilde Q$ are defined in \eqref{eq:cutoff} and \eqref{eq:eqvolumetilde}, respectively. The function $H(M)$ is the ex-ante trading profit per informed trader, net of execution costs and priority fees, and satisfies $\limsup_{M\to\infty} H(M)\le 0$. If $C>0$ and $H(2) > C$, then an equilibrium number $M^\star\ge2$ of informed traders exists and is given by the largest integer $M$ satisfying \eqref{eq:eqconditionM}. Moreover, $M^\star$ is non-decreasing in the noise-trading mass $N$,
and non-increasing in the information cost $C$.
\end{proposition}

In the equilibrium condition \eqref{eq:eqconditionM}, the integrand on the right-hand side represents the expected profit from participating in the PGA when a trader observes a valuation $v > \underline v_M$ and liquidity supply is set at its equilibrium level \eqref{eq:liq supp eq}. These expected profits are weighted by the likelihood $dF(v)$ of observing valuation $v$ in stage three. Proposition \ref{prop:eq nb traders} also shows that the mass of informed traders decreases with the cost of information and increases with the size of uninformed demand.

In markets, price efficiency is directly related to the number of traders who incur the cost of acquiring information. Each revealed valuation reduces the conditional variance of the fundamental price from the perspective of an uninformed observer. Thus, lower information costs $C$ and greater noise trading improve price efficiency. However, under paid-priority discriminatory pricing of blockchains,  price efficiency is also related to the participation cutoff \eqref{eq:cutoff} below which traders abstain from the PGA. An increase in entry raises the participation cutoff, so that only increasingly high valuations are revealed. Information becomes more truncated, and price efficiency deteriorates.

\section{Block time}\label{sec:blocktime}

Our model endogenizes trading volumes, entry into the blockchain, participation in the PGA, liquidity supply, and the amount of information in the market. The primitives are the price-sensitivity and total mass of noise trading flow, the distribution of private valuations, and the cost of acquiring information. An additional primitive specific to blockchains is the length of the slot, i.e., block time. 

Block time is the delay traders must wait before the market clears, and it is necessary to secure the blockchain. It therefore shapes both the distribution of private valuations at clearing and the intensity of uninformed trading within a block. Specifically, longer blocks increase the likelihood of large innovations in prices and widen the dispersion of private valuations, and they also deteriorate execution quality for uninformed traders within the block.

Let $T$ denote block time, which is common knowledge to all competitive agents. We make the following assumptions to formalize the effect of block time on the primitives of our model.

\begin{assume}\label{assume:T}
\begin{enumerate}
\item[(i)] The distribution $F(\cdot\,;T)$ of private valuations has support $[0,\overline v(T)]$ where $\overline v(T)$ is strictly increasing in block time $T$, with $\lim_{T\to\infty}\overline v(T)=\infty$. For any $T'>T$, $F(\cdot\,;T')$ strictly first-order stochastically dominates $F(\cdot\,;T)$ on $(0,\overline v(T)]$, i.e., $F(v;T')<F(v;T)$ for all $v\in(0,\overline v(T)]$.
\item[(ii)] The mass of noise trading flow $N(T)$ is decreasing in $T$ and nonnegative.
\end{enumerate}
\end{assume}

Assumption~\ref{assume:T}(i) states that longer block times widen the support of private valuations and make extreme valuations more likely. Assumption~\ref{assume:T}(ii) states that longer block times reduce the mass of noise trading flow. This reflects greater uncertainty about execution prices within longer blocks. The following corollary presents the comparative statics implied by these assumptions and the results derived in the previous sections.

\begin{corollary}\label{cor:blocktime}
Fix $M\ge 2$ and let Assumption~\ref{assume:T} hold. The participation cutoff $\underline v_M(T)$ in~\eqref{eq:cutoff} is increasing in $T$. With many informed traders ($M\to\infty$), the DEX price $\overline v(T)-\pi$ at the end of the block is increasing in $T$, the equilibrium liquidity
\(
\dfrac{8\,\pi\,N(T)}{3\,(\overline v(T)-\pi)^2}-\theta
\)
is decreasing in $T$, and markets shut down for sufficiently large $T$.
\end{corollary}

Corollary~\ref{cor:blocktime} shows that longer block times discourage participation in informed trading and push prices to more extreme levels, which impairs price discovery. The result also shows that, when many traders compete, longer blocks reduce liquidity, and beyond a critical block time, the market collapses. 

Longer block times reduce uninformed demand, by Assumption~\ref{assume:T}(ii). Holding the valuation distribution fixed, this channel lowers DEX liquidity for any fixed number of informed traders, tightens the market viability condition~\eqref{eq:profitability condition}, and reduces the  number of informed traders $M^\star$. 

In traditional markets with continuous trading, transactions that would be aggregated within a single blockchain slot of duration $T$ instead unfold over multiple short clearing rounds. Greater competition among informed traders in such markets accelerates the incorporation of information in early rounds, and market depth typically improves thereafter; see~\cite{holden1992long}. By contrast, blockchain markets aggregate the entire interval $T$ into a single trading round with a priority gas auction. Our results show that longer block times, although they enhance security, decrease liquidity, deter entry, and impair price efficiency.

\section{Conclusions}

The stated ambition of blockchains is to provide a decentralized alternative to intermediaries and to organize economic activity at scale. Financial markets are a central component of that ambition. 
We proposed a theoretical model of price formation and liquidity  in a blockchain market with discriminatory pricing based on paid-priority queues and discrete clearing. The model endogenized trading volumes, entry into informed trading, participation in the priority gas auction, liquidity supply, and payment for information. We showed that market outcomes are governed by three key primitives: the distribution of private valuations, the mass of uninformed trading flow, and the length of the blockchain slot. 

Our analysis showed that the very features that secure decentralization, most notably block time to conduct consensus and priority fees to compensate block builders, can undermine market viability. In general, under these design features, blockchains are not viable for price discovery.
If blockchains are to become a credible infrastructure for price discovery, protocol design must carefully address how transactions are organized and prioritized within each block.

\clearpage

\appendix

\section{Institutional details}\label{sec:instituionaldetails}

The microstructure of DEXs differs fundamentally from that
of traditional financial markets due to the unique
 infrastructure on which they operate.
In this section, we outline some institutional features of
blockchains that underpin the framework of our model.

\subsection{Ethereum, memory pools, and priority gas auctions}
\label{apx:pgas}

The Ethereum blockchain is a distributed and public digital ledger stored by \textit{nodes}.
Blocks of transactions are sequentially added to the ledger at regular intervals called
block time.
Block time must be long enough to allow for secure
confirmation of blocks because this process requires aggregating a large number
of cryptographic signatures in multiple rounds of network communications.
Shortening block time excludes slower participants and concentrates rewards
among those with superior connectivity, which would undermine decentralization.
For example, block time on Ethereum is deterministic and fixed to $12$-second.

When an agent submits a transaction,
it is not immediately included in a block.
Instead, the pending transaction is stored in a memory pool
until the end of block time.
For a  pending transaction to be executed,
it must be included in a block created by a builder.
Builders are paid gas fees by agents. Gas fees consist of a
mandatory base fee and an optional priority fee.
The priority fee incentivizes quick inclusion and priority in the block.
Thus, agents compete in PGAs for
better execution prices or to exploit arbitrage opportunities.

Accordingly, there are three stages in the lifecycle of a transaction:
(i) \emph{submission}, where an agent specifies the transaction details
and sets a gas fee;
(ii) \emph{storage in the memory pool}, where the
pending transaction awaits selection by a builder;
and (iii) \emph{inclusion in a block}, which corresponds
to the execution of the transaction.
Transaction submission may occur at any time within
the blockchain slot, and it is executed provided the
transaction reaches the builder's memory pool
before block creation.\footnote{In practice, if there are too many
transactions in the memory pool, only a fraction of all transactions is executed.}

\paragraph{Memory pools and PGAs.}
There are two types of memory pools: public and private.
Pending transactions in the public memory pool are visible prior to execution,
so agents operating on the Ethereum blockchain are exposed to arbitrage
and predatory attacks. Private memory pools have recently emerged in
the blockchain ecosystem to protect users;
see~\cite{capponi2024maximal, wang2025private}.\footnote{\href{https://www.flashbots.net/}{www.flashbots.net}}
Agents in these private channels send their transactions
directly to builders to conceal transaction details
and avoid attacks. At present, the vast majority of trading
flow in Ethereum DEXs passes through private memory
pools.\footnote{See
\href{https://dune.com/queries/5184076/8532375}{www.dune.com/queries/5184076/8532375}.}
For additional details on blockchain protocols, see~\cite{john2025economics}.

Next, we argue that our model is suitable for PGAs among
informed traders in both types of memory pools when the
blockchain is the sole venue for price formation.

PGAs differ across public and private memory pools. In private memory pools,
traders do not observe each other's bids before block creation,
so the PGA is sealed-bid as described in our model above.
In contrast, in public memory pools, bids are visible prior to execution,
and traders compete in an online auction throughout the blockchain
slot.\footnote{Another difference between both types of memory pools is that
in public memory pools, failed transactions
are paid regardless of execution, whereas in private memory pools,
transactions that are not executed do not pay priority fees.}

\begin{figure}[!h]
    \centering
    \includegraphics[width=\linewidth]{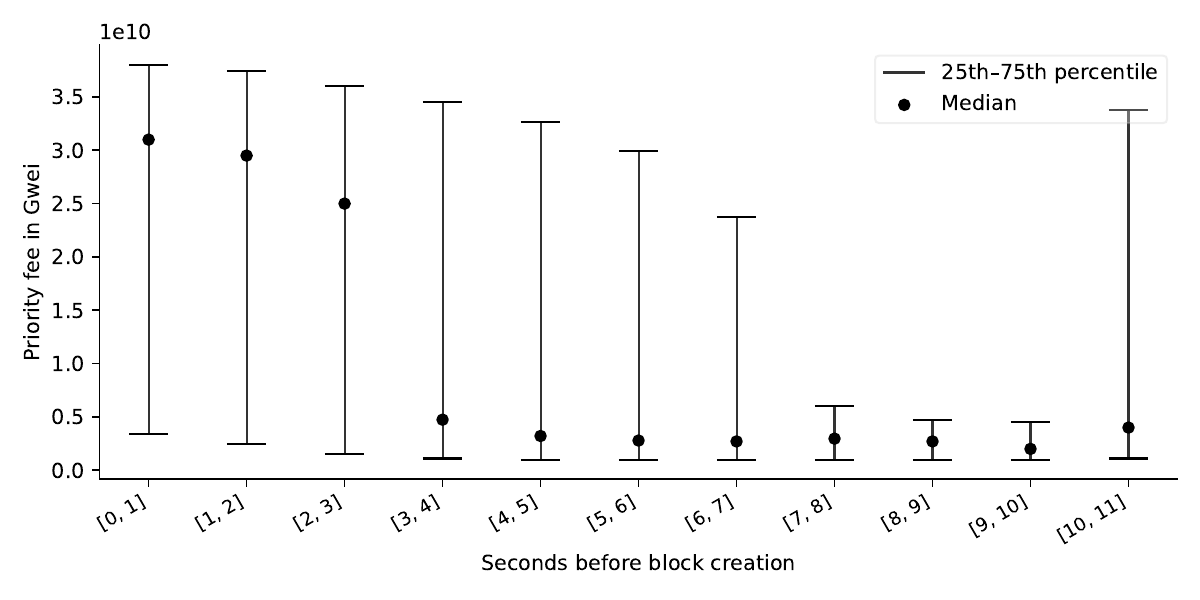}
    \caption{Distribution of priority fees as a function of
    timing within the blockchain slot.  Informed traders are
    typically associated with higher priority fees,
    as shown empirically in~\cite{capponi2023price} and
    theoretically in our model below. The figure shows
    that they tend to submit their transactions very
    close to the end of the blockchain slot. The data
    is from \textit{EthPandaOps} and it includes $10^7$
    transactions observed in the public memory pool of
    Ethereum between 20~March~2024, 00:00, and
    21~March~2024, 17:08. This data corresponds to a
    period when the public memory pool accounted for
    approximately $60\%$ of DEX trading volumes.
    At present (November 2025), this share is about $16\%$.
    The SQL data tables and other details
    are \href{https://github.com/ethpandaops/xatu-data}{here}.
    See  \cite{liu2022empirical, mizrach2025marginal} for a more
    detailed empirical analysis.}
    \label{fig:distribPF}
\end{figure}

In public memory pools, informed traders benefit from waiting
until the end of block time to submit
their transactions for three reasons. First, early bids reveal
information to competitors, second, early bids set a minimum fee due to the
\textit{account nonce};\footnote{A \emph{nonce} tracks the number of
transactions and it is assigned to each user to prevent replay and
double-spending attacks; see  \cite{vujivcic2018blockchain}.
Users can only modify pending transactions
by submitting a new one with the same  nonce and a higher gas fee.
}
and third, late bidding allows users to incorporate the most
up-to-date information. As a result, late bidding has
emerged as the dominant strategy in public memory pools.
We study the distribution of
transaction submission time in the public memory pool
of Ethereum in Figure~\ref{fig:distribPF} to support this claim empirically.
Therefore, our model is suitable to model PGAs in public memory pools.

\
\
At present, PGAs are mainly between specialized users (searchers or bots)
who compete to exploit the discrete-time structure of blockchains
through  two types of strategies: attacks on transactions
in public memory pools,\footnote{These include frontrunning,
backrunning, and sandwich attacks.} and
arbitrage opportunities across DEXs or
between a DEX and a competing centralized venue
such as Binance; see~\cite{heimbach2024non}.
Effectively, this type of PGA is a first-price
sealed-bid auction in which the value of
the item is public and known to all competitors.
As predicted by standard theory and as observed empirically,
the majority of the revenue from such PGAs is ultimately
paid back to builders in priority fees.
The objective of this paper is not to study PGAs for arbitrage.
Instead, we focus on market outcomes when PGAs
are the means for price formation
and private information is incomplete.

\paragraph{Sequential market clearing.}
In our model, we make the simplifying assumption that builders strictly
sort transactions in the block only by priority fee.  In practice, they are not
required by the blockchain protocol to do so.  However, failing to do so
harms reputation and weakens agents' incentive to pay priority fees.
Moreover, the Ethereum Foundation is actively improving the blockchain protocol to
impose rules for the behavior of builders to reduce censorship and rent
extraction; see \cite{thiery2024focil,wadhwa2025aucil}.

\subsection{Automated market makers} \label{apx:amms}

In this section, we show how the market frictions assumed in
our model above apply to automated market makers (AMMs).
In AMMs, algorithmic rules specify how liquidity supply,
in the form of aggregated reserves in a liquidity pool,
determines execution prices and price impact;
see~\cite{lehar2025decentralized}, \cite{capponi2021adoption},
and  \cite{cartea2025decentralised}.

Assume an AMM that
supplies liquidity $x$ in
the reference asset $X$ and $y$ in the risky asset $Y.$
A convex trading function $\Gamma$
defines the combinations of reserves $\{x=\Gamma(y),y\}$ that
make LPs indifferent, i.e., an iso-liquidity curve.
Let $Q > 0$ denote a quantity of $Y$ that a trader
wishes to buy (resp.~sell). The trader pays (resp.~receives)
$\Gamma(y-Q)-\Gamma(y)$ (resp.~$\Gamma(y+Q)-\Gamma(y)$) in asset $X$.
Thus, execution prices per unit of asset $Y$ are
\begin{equation}\label{eq:exec prices CFM}
    \text{price to sell }Q:\ \  \frac{\Gamma(y)-\Gamma(y+Q)}{Q}\,,\qquad
    \text{price to buy }Q:\ \  \frac{\Gamma(y-Q)-\Gamma(y)}{Q}\,.
\end{equation}
Moreover, execution prices for an infinitesimal
volume $Q\rightarrow0$ converge to $V=-\Gamma'(y),$ which
is referred to as the \emph{marginal price}. The difference between
execution and marginal prices are execution
costs, and they are determined by the reserves $y$.
The reserves also determine the impact $-\Gamma'(y\pm Q) - V$
of a trade on the marginal price.

Our model assumes a linear price schedule and a linear price update rule in the DEX.
These are nonlinear in an AMM.
Consider the following approximations of execution prices:
\begin{align}\label{eq:slippage}
\frac{\Gamma(y)-\Gamma(y+Q)}{Q}
                        \approx V - \frac{Q\, \Gamma''(y)}{2}\,,
\qquad
\frac{\Gamma(y-Q)-\Gamma(y)}{Q}
                    \approx V + \frac{Q\, \Gamma''(y)}{2}\,,
\end{align}
and  price impact:
\begin{align}\label{eq:impact}
-\Gamma'(y+Q) - V \approx -Q\, \Gamma''(y)\,, \qquad
-\Gamma'(y-Q) - V  \approx Q\, \Gamma''(y)\,.
\end{align}

\begin{figure}[!h]
\centerline{
\includegraphics[]{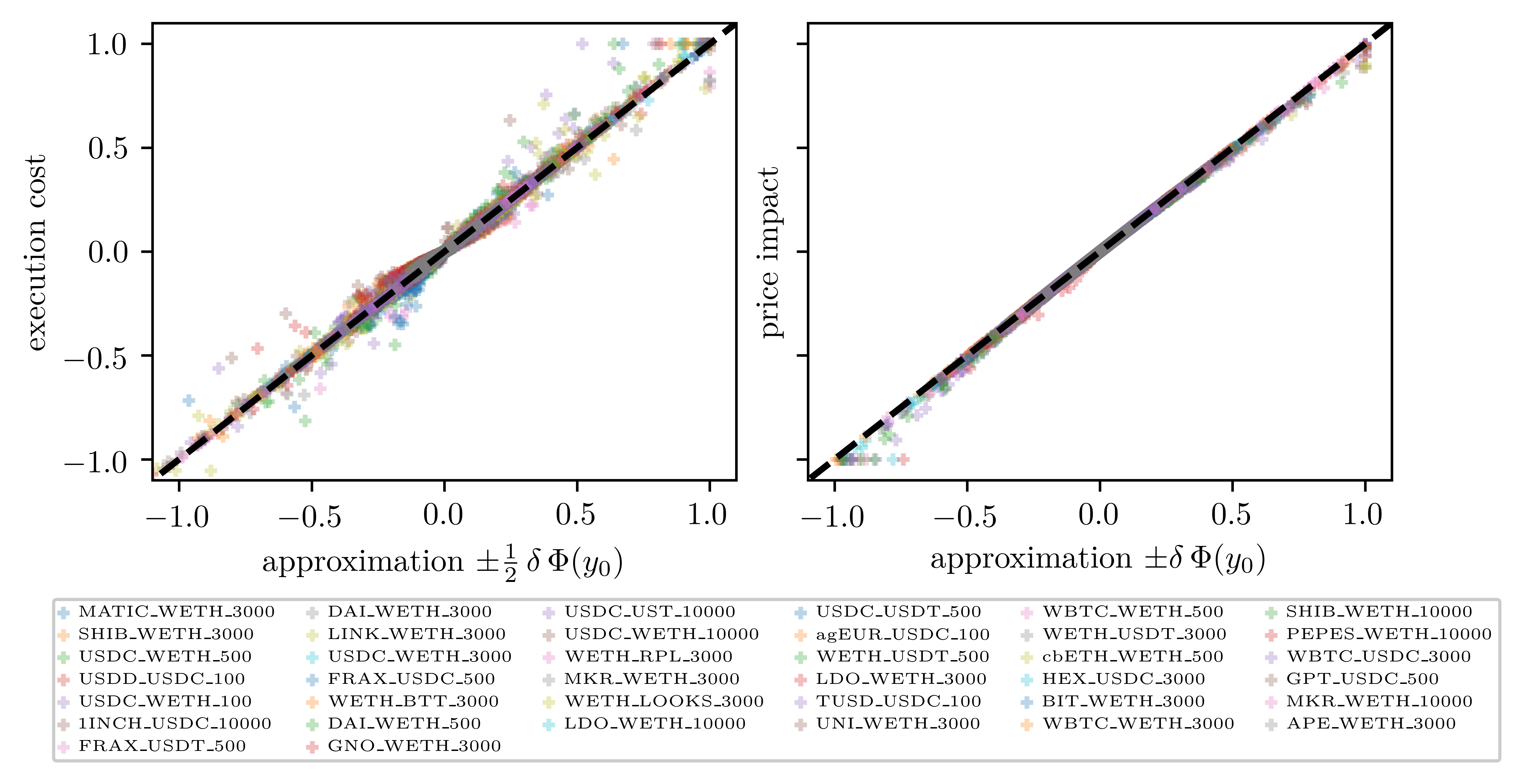}
}
  \caption{Scatter plots of the slippage and the price impact of $2.622$ million LT transactions against the approximations  \eqref{eq:slippage} and \eqref{eq:impact}. The transactions  are between $1$ January $2023$ and $31$ December $2023$ in $38$ different Uniswap v3 pools. For each pool, the slippages and price impacts are scaled between $[0,1]$ for buy orders and $ [-1,0]$ for sell orders. } \label{fig:approx}
\end{figure}

The depth $L$ of the DEX in our model depends on the convexity $\Gamma''(y)$
of the AMM's trading function in \eqref{eq:slippage} and \eqref{eq:impact}:\footnote{For example, the trading function for constant product markets like Uniswap, which are the most popular DEXs, is $\Gamma(y)=L^2/y$, where $L=\sqrt{x\,y}$ is the pool's liquidity parameter, so $\Gamma''(y) = 2\,V^{3/2}/L$, and by~\eqref{eq:sizepool} the model's depth is $L = 2/\Gamma''(y) = L/V^{3/2} = y^2/x$.}
\begin{equation}\label{eq:sizepool}
    L = 2 / \Gamma''(y).
\end{equation}
The convexity of trading functions is a key determinant of
liquidity; see \cite{milionis2022automated,cartea2023predictable}.
Akin to traditional electronic markets,
market frictions in AMMs increase with the traded volume $Q$
and decrease with the depth $L$.
Figure \ref{fig:approx} uses DEX data from the Ethereum blockchain
to show that approximating the AMM's pricing
rule with a linear schedule
is accurate in practice because trading
volumes represent a small fraction of total liquidity supply.

\
\

\clearpage

\section{Proofs}\label{apx:proofs}

\subsection{Proof of Lemma \ref{lem:simplifiedExpectedImpact}}
Fix trader~$i$. Since $G$ is continuous on $(0,\overline{Q}]$, the events $\{Q_{\left(j\right)}<\Phi^{-1}\left(\Phi_{i}\right)<Q_{\left(j+1\right)}\}$ for $j=0,\ldots,M-1$ form an almost-sure partition of the sample space, with the boundary cases understood as $\{\Phi^{-1}\left(\Phi_{i}\right)<Q_{\left(1\right)}\}$ for $j=0$ and $\{Q_{\left(M-1\right)}<\Phi^{-1}\left(\Phi_{i}\right)\}$ for $j=M-1$.

Whenever $Q_{\left(j\right)}<\Phi^{-1}\left(\Phi_{i}\right)<Q_{\left(j+1\right)}$, exactly $j$ of the $M-1$ competitor volumes fall below $\Phi^{-1}\left(\Phi_{i}\right)$, and these are precisely the volumes entering the lower partial sum $\Delta_{\left(1:j\right)}$. Therefore, on this event,
$$
\Delta_{\left(1:j\right)}=\sum_{k=1}^{M-1}Q_{k}\,\mathbbm{1}_{\{Q_{k}<\Phi^{-1}\left(\Phi_{i}\right)\}}\,,
$$
and the right-hand side does not depend on $j$. Multiplying by the corresponding indicator and summing over $j$, the partition property collapses the left-hand side:
$$
\sum_{j=0}^{M-1}\mathbbm{1}_{\{Q_{\left(j\right)}<\Phi^{-1}\left(\Phi_{i}\right)<Q_{\left(j+1\right)}\}}\,\Delta_{\left(1:j\right)}=\sum_{k=1}^{M-1}Q_{k}\,\mathbbm{1}_{\{Q_{k}<\Phi^{-1}\left(\Phi_{i}\right)\}}\,.
$$
Taking expectations and using that $Q_{1},\ldots,Q_{M-1}$ are i.i.d.\ with distribution $G$, we obtain
\begin{align*}
\sum_{j=0}^{M-1}\mathbb{E}_{i}\left[\mathbbm{1}_{\Phi_{\left(j\right)}<\Phi_{i}<\Phi_{\left(j+1\right)}}\Delta_{\left(1:j\right)}\right]&=\left(M-1\right)\,\mathbb{E}\left[Q\,\mathbbm{1}_{\{Q<\Phi^{-1}\left(\Phi_{i}\right)\}}\right]\\&=\left(M-1\right)\int_{0}^{\Phi^{-1}\left(\Phi_{i}\right)}x\,dG\left(x\right)\,.
\end{align*}

\qed

\subsection{Proof of Proposition \ref{prop:pf}}
\noindent First, consider the function
$$
\mathbb{E}\left[Q\mid Q>x\right]=\tau\left(x\right)\,,
$$
for the random variable $Q$ with continuous distribution $g$. Write
$$
\tau\left(x\right)=\frac{\int_{x}^{\overline{Q}}u\,g\left(u\right)\,du}{1-G\left(x\right)}\,,
$$
to obtain
\begin{align*}
\tau'\left(x\right)=&\frac{-x\,g\left(x\right)}{1-G\left(x\right)}+g\left(x\right)\frac{\int_{x}^{\overline{Q}}u\,g\left(u\right)\,du}{\left(1-G\left(x\right)\right)^{2}}\\=&\frac{g\left(x\right)}{1-G\left(x\right)}\left(\tau\left(x\right)-x\right)\,.
\end{align*}

\noindent
Thus, the FOC derived from the optimization problem in  \eqref{eq:optimisation1} is
\begin{align*}
\Phi^{'}\left(\Phi^{-1}\left(\Phi_{i}\right)\right)=&\left(2\,Q_{i}\Big/L\right)\,\left(M-1\right)g\left(\Phi^{-1}\left(\Phi_{i}\right)\right)\tau\left(\Phi^{-1}\left(\Phi_{i}\right)\right)\\&-\left(2\,Q_{i}\Big/L\right)\,\left(M-1\right)\left(1-G\left(\Phi^{-1}\left(\Phi_{i}\right)\right)\right)\tau'\left(\Phi^{-1}\left(\Phi_{i}\right)\right)\\
=&\frac{2\,Q_{i}}{L}\,\left(M-1\right)g\left(\Phi^{-1}\left(\Phi_{i}\right)\right)\mathbb{E}\left[Q\mid Q>\Phi^{-1}\left(\Phi_{i}\right)\right]\\&-\frac{2\,Q_{i}}{L}\,\left(M-1\right)g\left(\Phi^{-1}\left(\Phi_{i}\right)\right)\left(\mathbb{E}\left[Q\mid Q>\Phi^{-1}\left(\Phi_{i}\right)\right]-\Phi^{-1}\left(\Phi_{i}\right)\right)\\=&\frac{2\,Q_i}{L}\left(M-1\right)\,\Phi^{-1}\left(\Phi_{i}\right)\,g\left(\Phi^{-1}\left(\Phi_{i}\right)\right)\,.
\end{align*}
In equilibrium, trader $i$ finds it optimal to adopt the same strategy $\Phi$, so $\Phi^{-1}\left(\Phi_{i}\right) = Q_i$, so the FOC becomes
$$
\Phi^{'}\left(Q_{i}\right)=\frac{2}{L}\left(M-1\right)\,Q_{i}^{2}\,g\left(Q_{i}\right),
$$
with boundary condition $\Phi(0) = 0$, which establishes~\eqref{eq:priority fee eq model 1}. Substituting $\Phi_i=\Phi(Q_i)$ into~\eqref{eq:optimisation1} and applying Lemma~\ref{lem:simplifiedExpectedImpact} with $\Phi^{-1}(\Phi_i)=Q_i$ yields
$$
-\Phi(Q_i)+\frac{2\,Q_i}{L}(M-1)\int_0^{Q_i} x\,dG(x)
= \frac{2}{L}(M-1)\int_0^{Q_i} x(Q_i-x)\,dG(x),
$$
which is~\eqref{eq:eqobjpf}.

\noindent It remains to verify global optimality. Reparametrizing the objective \eqref{eq:optimisation1} in the bid $x^\star=\Phi^{-1}(\Phi_i)$ gives $U(x^\star)=-\Phi(x^\star)+\frac{2Q_i}{L}(M-1)\int_0^{x^\star}x\,dG(x)$. Using the closed form $\Phi'(y)=(2/L)(M-1)\,y^2\,g(y)$ just derived,
$$
U'(x^\star) \;=\; -\Phi'(x^\star) + \frac{2Q_i}{L}(M-1)\,x^\star g(x^\star) \;=\; \frac{2(M-1)}{L}\,x^\star\,g(x^\star)\,(Q_i - x^\star)\,,
$$
which is strictly positive on $(0,Q_i)$ and strictly negative on $(Q_i,\overline{Q}]$ whenever $g>0$. Bidding $\Phi_i>\Phi(\overline{Q})$ wins against all competitors with probability one and yields payoff $-\Phi_i+\frac{2Q_i}{L}(M-1)\int_0^{\overline{Q}}x\,dG(x)$, strictly decreasing in $\Phi_i$, so it is dominated by $\Phi_i=\Phi(\overline{Q})$. Hence $x^\star=Q_i$ is the unique global maximizer and the FOC's solution characterizes the global optimum.
\qed

\subsection{Proof of Lemma \ref{lem:cutoff}}

For $M\ge 2$ and $v\in[0,\overline v]$, define the function
\begin{equation}\label{eq:proof:lem2}
h_M(v)= v-\pi-\int_{v}^{\overline v}\bigl(e^{(1-F(u))(M-1)}-1\bigr)\,du,
\qquad v\in[0,\overline v].
\end{equation}

\noindent
First, we show that the function $h_M$ is increasing in $v.$ Note that the integrand $e^{(1-F(u))(M-1)}-1$ is continuous and nonnegative. Thus, $h_M$ is continuous and differentiable on
$(0,\overline v)$ with derivative
\[
h_M'(v)=
e^{(1-F(v))(M-1)}>1.
\]
Thus $h_M$ is strictly increasing.

\noindent
Next we show the existence and uniqueness of the root. Because $h_M$ is strictly increasing, it can have at most one root. Use the endpoint conditions
\[
h_M(0) < 0
\qquad\text{and}\qquad
h_M(\overline v) = \overline v - \pi > 0\,.
\]
By continuity of $h_M$, we obtain existence of a unique $\underline v_M\in(0,\overline v)$
satisfying $h_M(\underline v_M)=0$.  Since $h_M(\overline v)>0$, the root cannot equal $\overline v$ and thus $\underline v_M<\overline v$.

\noindent
Next, we show that the root, i.e., the valuation cutoff, is increasing in the number of informed traders $M$ that paid the cost of acquiring information in stage one.  For fixed $v$, differentiate $h_M(v)$ with respect to $M$ to obtain:
\[
\partial_Mh_M(v)
= -\int_v^{\overline v} (1-F(u))\,e^{(1-F(u))(M-1)}\,du < 0.
\]
Thus for every $v$, the function $M\mapsto h_M(v)$ is strictly decreasing. Use the identity
$h_M(\underline v_M)=0$ and the implicit function theorem{\footnote{Although $M$ takes integer values, $h_M(v)$ is well-defined for any real $M\ge 2$, so the implicit function theorem applies. Equivalently, for integer $M$, $h_{M+1}(\underline v_M)<h_M(\underline v_M)=0$ because the integrand in~\eqref{eq:proof:lem2} is strictly increasing in $M$ on the set where $F<1$; combined with strict increasingness of $h_{M+1}$ in $v$, this yields $\underline v_{M+1}>\underline v_M$.}} to write
\[
\partial_M \underline v_M
= -\frac{\partial_M h_M(\underline v_M)}{\partial _v h_M(\underline v_M)}
> 0,
\]
because $\partial_M h_M(\underline v_M) < 0$ and $h_M'(\underline v_M) > 0$.
Therefore, the cutoff $\underline v_M$ is strictly increasing in $M$.

\noindent
Finally, we show that $\lim_{M\rightarrow\infty}\underline v_M = \overline v.$ First, we use \eqref{eq:proof:lem2} to obtain that for all $v<\overline v$, $\lim_{M\rightarrow\infty} h_M(v) = -\infty$.\footnote{For $u\in[v,\overline v)$, $1-F(u)>0$, so $e^{(1-F(u))(M-1)}\to+\infty$ pointwise and the integrand in~\eqref{eq:proof:lem2} diverges to $+\infty$ on a set of positive Lebesgue measure. By monotone convergence, $\int_v^{\overline v}\bigl(e^{(1-F(u))(M-1)}-1\bigr)\,du\to+\infty$ as $M\to\infty$, hence $h_M(v)\to-\infty$.} Thus, for each $\epsilon>0$, there exists a large enough $M(\epsilon)$ such that for all $M>M(\epsilon)$, $h_M(\overline v-\epsilon)<0.$ Use that (i) $h_M(\cdot)$ is increasing, (ii) $h_M(\overline v-\epsilon)<0$, and (iii) the definition of the cutoff $h_M(\underline v_M) = 0$, to obtain that for all $\epsilon>0$ and $M>M(\epsilon)$, $\overline v-\epsilon \le \underline v_M$. Next, we showed earlier that for all $M,$ $\underline v_M \le \overline v.$ Thus, for all $M > M(\epsilon)$,
$$
\overline v - \epsilon \le \underline v_M \le \overline v,
$$
and we conclude that $\lim_{M\rightarrow\infty}\underline v_M = \overline v.$

\qed
\subsection{Proof of Proposition \ref{prop:volume}}

\noindent Use the optimal priority fee \eqref{eq:priority fee eq model 1} in the objective \eqref{eq:optimisationvolumes} to write the expected wealth of trader $i$ as
$$
\mathbb{E}_{i}\left[W_{i}\right]=-\frac{2}{L}\left(M-1\right)\left(\int_{0}^{Q_{i}}x^{2}\,dG\left(x\right)+Q_{i}\int_{Q_{i}}^{\overline{Q}}x\,dG\left(x\right)\right)+Q_{i}\,\left(v_{i}-\pi-\frac{Q_{i}}{L}\right)-C\,.
$$

\noindent The optimal trading volume is the solution of $\partial_{Q_i} \mathbb{E}\left[W_{i}\right]= 0$ or
\begin{equation}\label{eq:proof volume eq1}
    0=\left(v_{i}-\pi\right)L-2\,Q_{i}-2\,\left(M-1\right)\int_{Q_{i}}^{\overline{Q}}x\,dG\left(x\right)\,.
\end{equation}
Differentiating \eqref{eq:proof volume eq1} in $Q_i$ yields the second-order condition
$$
\partial^2_{Q_i}\mathbb{E}_{i}\left[W_{i}\right] \;=\; \frac{2}{L}\bigl[(M-1)\,Q_i\,g(Q_i) - 1\bigr]\,,
$$
where the $Q_i^2 g(Q_i)$ contributions from the two integrals in the objective cancel. Under the equilibrium relation $G(Q(v))=F(v)$, the strategy $Q:[\underline v_M,\overline v]\to[0,\overline{Q}]$ is a strict bijection (established below), so every $Q_i\in(0,\overline{Q}]$ admits a unique $v=Q^{-1}(Q_i)$. Differentiation then gives $g(Q_i)=f(v)/Q'(v)$, and the ODE established below implies
$$
(M-1)\,Q_i\,g(Q_i) \;=\; \frac{(M-1)\,Q(v)\,f(v)}{L/2 + (M-1)\,Q(v)\,f(v)} \;<\; 1\,,
$$
so $\partial^2_{Q_i}\mathbb{E}_{i}\left[W_{i}\right]<0$ for every $Q_i\in(0,\overline{Q}]$, and also at $Q_i=0$ since the factor $Q_i g(Q_i)$ vanishes there. Hence the objective is strictly concave on $[0,\overline{Q}]$. For $Q_i>\overline{Q}$, $g$ vanishes and $\partial_{Q_i}\mathbb{E}_{i}\left[W_{i}\right]=v_i-\pi-2Q_i/L\le v_i-\overline v\le 0$, so the objective is strictly decreasing on $[\overline{Q},\infty)$. Combining, the FOC's solution is the unique global maximizer.

Recall that, in equilibrium, $G$ is the distribution of trading volumes pinned down by the equilibrium increasing trading volume function $Q$, so
$$
G\left(Q_{i}\right)=F\left(v_{i}\right)\,.
$$
Rearranging \eqref{eq:proof volume eq1} one obtains the boundary condition
$$
\overline{Q}=L\frac{\overline{v}-\pi}{2}
$$
and differentiating we find that the trading volumes solve the ODE
$$
Q'\left(v_{i}\right)=\frac{L}{2}+\left(M-1\right)Q\left(v_{i}\right)\,f\left(v_{i}\right)\,,
$$
whose solution is in the statement of the result. The solution is increasing in $v$, and trader $i$'s optimization problem is over nonnegative volumes. Thus the boundary condition is at the lowest valuation $\underline v_M$ for which the trader optimally chooses a nonnegative volume. This cutoff is characterized in Lemma \ref{lem:cutoff} and corresponds to when the FOC \eqref{eq:proof volume eq1}  is satisfied at $\underline v_M$ with $Q(\underline v_M) = 0$. For $v_{i}<\underline v_M$, evaluating the partial derivative of the objective at $Q_{i}=0$ yields
$$
\partial_{Q_{i}}\mathbb{E}_{i}\left[W_{i}\right]\Big|_{Q_{i}=0} = \left(v_{i}-\pi\right)L - 2\left(M-1\right)\int_{0}^{\overline{Q}} x\,dG\left(x\right) = \left(v_{i} - \underline v_{M}\right)L < 0\,,
$$
where the second equality uses \eqref{eq:cutoff} rewritten as the FOC at $Q_{i}=0$, $v_{i} = \underline v_{M}$. By strict concavity of the objective on $[0,\overline{Q}]$, the constrained maximum on $Q_{i}\ge 0$ is at $Q_{i} = 0$, hence $Q(v_{i}) = 0$. To obtain the form in Lemma \ref{lem:cutoff}, set $\tilde{Q}(\underline v_M)=0$ in \eqref{eq:eqvolumetilde}, which yields
$$
(\overline v-\pi)\,e^{-(M-1)(1-F(\underline v_M))} \;=\; \int_{\underline v_M}^{\overline v} e^{-(M-1)(F(u)-F(\underline v_M))}\,du\,.
$$
Multiplying both sides by $e^{(M-1)(1-F(\underline v_M))}$ gives
$$
\overline v-\pi \;=\; \int_{\underline v_M}^{\overline v} e^{(M-1)(1-F(u))}\,du\,,
$$
which, after adding and subtracting $\overline v-\underline v_M$, is equivalent to \eqref{eq:cutoff}.

It remains to establish the monotonicity and limit properties in the statement. From the ODE, $Q'(v_i) = L/2 + (M-1)Q(v_i)\,f(v_i) \ge L/2 > 0$ on $[\underline v_M,\overline v]$, so $Q$ is strictly increasing in $v_i$. Next, \eqref{eq:eqvolumetilde} rewrites as
$$
\tilde{Q}(v_i) = \frac{e^{-(M-1)(1-F(v_i))}}{2}\left[(\overline v - \pi) - \int_{v_i}^{\overline v} e^{(M-1)(1-F(u))}\,du\right],
$$
where the prefactor is positive and decreasing in $M$ since $1-F(v_i)\ge 0$. The bracket is decreasing in $M$ since its integrand increases in $M$, and nonnegative on $[\underline v_M,\overline v]$ since it vanishes at $v_i=\underline v_M$ by \eqref{eq:cutoff} and is increasing in $v_i$ (its integrand is positive). Hence $Q(v_i)=L\,\tilde{Q}(v_i)$ is decreasing in $M$. Finally, bounding the bracket above by $\overline v-\pi$ yields
$$
\tilde{Q}(v_i) \;\le\; \frac{\overline v-\pi}{2}\,e^{-(M-1)(1-F(v_i))},
$$
which converges to zero as $M\to\infty$ for every $v_i<\overline v$.
\qed

\subsection{Proof of Proposition \ref{prop:eqpfwithvol}}

Trader $i$ uses the priority fee \eqref{eq:priority fee eq model 1} for an arbitrary distribution of volumes with cumulative distribution function $G$:
\begin{equation}
\Phi\left(Q_{i}\right)=\frac{2}{L}\left(M-1\right)\int_{0}^{Q_{i}}u^{2}\,dG\left(u\right)\,.
\end{equation}

\noindent
The optimal volumes of PGA competitors are distributed according to the distribution $G$ described in \eqref{eq:distGeq}:
\begin{equation}
G\left(\{0\}\right) = F(\underline v_M)
\qquad \text{and} \qquad
G\big(Q(v)\big) = F(v)
\quad \text{for } v \in [\underline v_M, \overline v]\,,
\end{equation}

\noindent
The priority fee is trivially zero for $Q_i=0$, i.e., when $v_i\le \underline v_M.$ Next, fix $v_i\in [\underline v_M,\overline v].$ Note that the atom at zero causes no problem because the integrand $u^2$ evaluated at zero is zero. Thus, the mass $G(\{0\})$ contributes zero to the Stieltjes integral, and we write
\begin{equation}
\int_0^{Q_i}  u^2\,dG(u) = 0\,G(\{0\}) + \int_{(0,Q_i]}u^2\,dG(u) = \int_{\underline v_M}^{v_i}Q(x)^2\,dF(x)\,.
\end{equation}
where we used that $Q:[\underline v_M,\overline v] \mapsto [0,Q(\overline v)]$, defined in \eqref{eq:eqvolume}, is increasing and continuous, hence it is invertible on its image and $G(Q(v))=F(v)$ for $v\ge \underline v_M.$ Substituting $Q(x) = L\,\tilde{Q}(x)$, so that $Q(x)^2 = L^2\,\tilde{Q}(x)^2$, and combining with the prefactor $2/L$ from \eqref{eq:priority fee eq model 1}, we obtain \eqref{eq:eqpfwithvol}.

\noindent
Since $\tilde{Q}$ in \eqref{eq:eqvolumetilde} does not depend on $L$, $\Phi(v_i)$ is linear in $L$ by \eqref{eq:eqpfwithvol}, hence increasing in $L$.

Next, we show that for all $v_i<\overline v$, the priority fee tends to zero when the number of competitors $M$ tends to infinity. For $v_i \le \underline v_M$, $\Phi(v_i) = 0$ by the first part of the proposition. We therefore consider $v_i \in (\underline v_M, \overline v)$ in what follows.
Consider the strategic volumes \eqref{eq:eqvolumetilde}:
\[
\tilde{Q}(x)
=
\frac{1}{2}
\Big(
(\overline v-\pi)e^{-(M-1)(1-F(x))}
-
\int_x^{\overline v}
e^{-(M-1)(F(u)-F(x))}\,du
\Big) \quad \text{if }x\ge\underline v_M,
\]
and zero otherwise. By Proposition~\ref{prop:volume}, $\tilde{Q}(x)\ge 0$ on $[\underline v_M,\overline v]$, and the second term in $\tilde{Q}(x)$ is nonnegative, so
\[
0 \le \tilde{Q}(x)
\le
\frac{\overline v-\pi}{2}
\,e^{-(M-1)(1-F(x))}
\implies
\tilde{Q}(x)^2
\le
\frac{(\overline v-\pi)^2}{4}
e^{-2(M-1)(1-F(x))}.
\]

\noindent
Fix $v_i < \overline v$. Since $F$ is nondecreasing, for all $x \le v_i$ we have $F(x) \le F(v_i)$ and therefore
\[
1-F(x) \ge 1-F(v_i) > 0.
\]
Thus,
\[
\tilde{Q}(x)^2
\le
\frac{(\overline v-\pi)^2}{4}
e^{-2(M-1)(1-F(v_i))}
\quad
\text{for all } x \in [\underline v_M, v_i].
\]

\noindent
It follows from the definition of the equilibrium priority fee in \eqref{eq:eqpfwithvol}, accounting for strategic volumes, that
\begin{equation}\label{eq:proof prop 3 1}
\Phi(v_i)=
2L (M-1)
\int_{\underline v_M}^{v_i}
\tilde{Q}(x)^2 \, dF(x)
\le
2L (M-1)
(F(v_i)-F(\underline v_M))
\frac{(\overline v-\pi)^2}{4}
e^{-2(M-1)(1-F(v_i))}.
\end{equation}
Since $F(v_i)-F(\underline v_M) \le 1$, we obtain
\[
0 \le
\Phi(v_i)
\le
\frac{L (\overline v-\pi)^2}{2}
(M-1)
e^{-2(M-1)(1-F(v_i))}.
\]
Because $1-F(v_i) > 0$ for all $v_i < \overline v$,
we conclude that, for all $v_i<\overline v$,
\[
\lim_{M \to \infty}
\Phi(v_i) = 0.
\]
\qed

\subsection{Proof of Proposition \ref{prop:activetraders}}

\noindent
We first establish the closed form \eqref{eq:aggvolclosedform}.  The first-order condition \eqref{eq:proof volume eq1} characterising the optimal trading volume reads
\[
0 \;=\; (v_i-\pi)\,L - 2\,Q_i - 2\,(M-1)\int_{Q_i}^{\overline{Q}}x\,dG(x).
\]
Evaluating this condition at $v_i=\underline v_M$, and using $Q(\underline v_M)=0$ from Proposition~\ref{prop:volume}, yields
\[
(\underline v_M-\pi)\,L \;=\; 2\,(M-1)\int_{0}^{\overline{Q}}x\,dG(x).
\]
By \eqref{eq:distGeq} and \eqref{eq:eqvolume}, the atom $G(\{0\})=F(\underline v_M)$ contributes zero to the Stieltjes integral (the integrand $x$ vanishes at $x=0$), and the change of variable $x=Q(v)=L\,\tilde{Q}(v)$ on $[\underline v_M,\overline v]$ gives
\[
\int_{0}^{\overline{Q}}x\,dG(x) \;=\; \int_{\underline v_M}^{\overline v}Q(v)\,dF(v) \;=\; L\int_{\underline v_M}^{\overline v}\tilde{Q}(v)\,dF(v).
\]
Combining the two displays,
\[
(\underline v_M-\pi)\,L \;=\; 2\,(M-1)\,L\int_{\underline v_M}^{\overline v}\tilde{Q}(v)\,dF(v),
\]
and multiplying by $M/(2(M-1))$ yields \eqref{eq:aggvolclosedform}.  The limit $L(\overline v-\pi)/2$ then follows directly from \eqref{eq:aggvolclosedform} and Lemma~\ref{lem:cutoff}, which gives $\lim_{M\to\infty}\underline v_M=\overline v$.  We present below an alternative direct derivation of this limit, which also establishes the auxiliary result $\lim_{M\to\infty}(M-1)(1-F(\underline v_M))=\infty$ used in the proof of Proposition~\ref{prop:liquidity supply}.

\noindent
Recall that for $v_i\ge\underline v_M$,
\[
\tilde{Q}(v_i)
=
\frac{1}{2}
\Bigg(
(\overline v - \pi)e^{-(M-1)(1-F(v_i))}
-
\int_{v_i}^{\overline v}
e^{-(M-1)(F(u)-F(v_i))}\,du
\Bigg).
\]
Factoring out $e^{-(M-1)(1-F(v_i))}$ yields
\[
\tilde{Q}(v_i)
=
\frac12
e^{-(M-1)(1-F(v_i))}
\left(
\overline v - \pi
-
\int_{v_i}^{\overline v}
e^{(M-1)(1-F(u))}\,du
\right).
\]

\medskip
\noindent
Hence, the expected total trading volume of informed traders is
\begin{align}
M \int_{\underline{v}_M}^{\overline{v}} \tilde{Q}(x)\,dF(x)
&=
\frac{M(\overline v-\pi)}{2}
\int_{\underline{v}_M}^{\overline{v}}
e^{-(M-1)(1-F(x))}\,dF(x)
\label{eq:term1}
\\
&\quad
-
\frac{M}{2}
\int_{\underline{v}_M}^{\overline{v}}
e^{-(M-1)(1-F(x))}
\left[
\int_{x}^{\overline{v}}
e^{(M-1)(1-F(u))}\,du
\right]
dF(x).
\label{eq:term2}
\end{align}

\medskip
\noindent
We first evaluate \eqref{eq:term1}. Using the change of variable $z=1-F(x)$,
\begin{align}
\frac{M(\overline v-\pi)}{2}
\int_{\underline{v}_M}^{\overline{v}}
e^{-(M-1)(1-F(x))}\,dF(x)
&=
\frac{M(\overline v-\pi)}{2}
\int_{0}^{1-F(\underline v_M)}
e^{-(M-1)z}\,dz
\nonumber
\\
&=
\frac{M(\overline v-\pi)}{2(M-1)}
\left(
1-e^{-(M-1)(1-F(\underline v_M))}
\right).
\label{eq:term1_co}
\end{align}

\medskip
\noindent
We now show that
\[
\lim_{M\to\infty}(M-1)\big(1-F(\underline v_M)\big)=\infty.
\]
Suppose instead that this sequence is bounded. Then there exists $C>0$ such that
\[
(M-1)\big(1-F(\underline v_M)\big)<C
\qquad\text{for all }M.
\]
Using the definition of $\underline v_M$ in \eqref{eq:cutoff},
\[
\underline v_M-\pi
=
\int_{\underline v_M}^{\overline v}
\bigl(e^{(M-1)(1-F(u))}-1\bigr)\,du,
\]
and the bound $1-F(u)\le 1-F(\underline v_M)$, we obtain
\[
0
\le
\underline v_M-\pi
\le
\bigl(e^{(M-1)(1-F(\underline v_M))}-1\bigr)
(\overline v-\underline v_M)
\le
\bigl(e^C-1\bigr)
(\overline v-\underline v_M).
\]
By Lemma \ref{lem:cutoff}, $\underline v_M\to\overline v$ as $M\to\infty$. The inequality above would then imply $\underline v_M\to\pi$, which contradicts Lemma \ref{lem:cutoff}. Hence,
\[
(M-1)\big(1-F(\underline v_M)\big)\to\infty.
\]

\medskip
\noindent
Combining this result with \eqref{eq:term1_co} gives
\[
\lim_{M\to\infty}
\frac{M(\overline v-\pi)}{2}
\int_{\underline{v}_M}^{\overline{v}}
e^{-(M-1)(1-F(x))}\,dF(x)
=
\frac{\overline v-\pi}{2}.
\]

\medskip
\noindent
We now turn to \eqref{eq:term2}. Define
\begin{align*}
I_2
&=
\frac{M}{2}
\int_{\underline{v}_M}^{\overline{v}}
e^{-(M-1)(1-F(x))}
\left[
\int_{x}^{\overline{v}}
e^{(M-1)(1-F(u))}\,du
\right]
dF(x).
\end{align*}
Reversing the order of integration,
\begin{align*}
I_2
&=
\frac{M}{2}
\int_{\underline{v}_M}^{\overline{v}}
e^{(M-1)(1-F(u))}
\left[
\int_{\underline v_M}^{u}
e^{-(M-1)(1-F(x))}\,dF(x)
\right]
du.
\end{align*}
With the change of variable $z=1-F(x)$,
\begin{align*}
\int_{\underline{v}_M}^{u}
e^{-(M-1)(1-F(x))}\,dF(x)
&=
\int_{1-F(u)}^{1-F(\underline v_M)}
e^{-(M-1)z}\,dz
\\
&=
\frac{1}{M-1}
\left(
e^{-(M-1)(1-F(u))}
-
e^{-(M-1)(1-F(\underline v_M))}
\right).
\end{align*}
Substituting back,
\[
I_2
=
\frac{M}{2(M-1)}
\int_{\underline{v}_M}^{\overline{v}}
\left(
1
-
e^{-(M-1)(1-F(\underline v_M))}
e^{(M-1)(1-F(u))}
\right)
du.
\]
Hence,
\[
I_2
=
\frac{M}{2(M-1)}
\left[
(\overline v-\underline v_M)
-
e^{-(M-1)(1-F(\underline v_M))}
\int_{\underline{v}_M}^{\overline{v}}
e^{(M-1)(1-F(u))}\,du
\right].
\]

\medskip
\noindent
Rearranging the cutoff equation \eqref{eq:cutoff} yields the exact identity
\[
\int_{\underline{v}_M}^{\overline{v}}
e^{(M-1)(1-F(u))}\,du
\;=\;
\overline v-\pi.
\]
Since $(M-1)(1-F(\underline v_M))\to\infty$, the exponential term vanishes, and therefore $I_2\to 0$ as $M\to\infty$.

\medskip
\noindent
Combining the limits of \eqref{eq:term1} and \eqref{eq:term2}, we conclude that
\[
\lim_{M\to\infty}
M
\int_{\underline v_M}^{\overline v}
\tilde{Q}(u)\,dF(u)
=
\frac{\overline v-\pi}{2}.
\]

\medskip
\noindent
Next, we show that the aggregate trading volume approaches its limit from below; that is,
\begin{equation}\label{eq:aggvolsaturation}
L\,M\int_{\underline v_M}^{\overline v}\tilde{Q}(x)\,dF(x) \;<\; L\,\frac{\overline v-\pi}{2}
\end{equation}
for all $M$ sufficiently large. By our standing assumption on $F$, there exists $\eta>0$ such that $f(u)\le 1/(\overline v-\pi)$ for all $u\in[\overline v-\eta,\overline v]$. We establish \eqref{eq:aggvolsaturation} by comparison with the uniform distribution on $[\pi,\overline v]$.

Let $F_U$ denote the uniform CDF on $[\pi,\overline v]$, $F_U(u)=(u-\pi)/(\overline v-\pi)$, and let $\underline v_M^{U}$ denote the corresponding cutoff defined by \eqref{eq:cutoff} with $F$ replaced by $F_U$.

\medskip
\noindent
We first compute the uniform cutoff and the corresponding uniform aggregate volume. Substituting $F_U$ in \eqref{eq:cutoff} and using the change of variable $s=(\overline v-u)/(\overline v-\pi)$,
\[
\underline v_M^{U}-\pi
\;=\;
\int_{\underline v_M^{U}}^{\overline v}\!\big(e^{(M-1)(\overline v-u)/(\overline v-\pi)}-1\big)\,du
\;=\;
\frac{\overline v-\pi}{M-1}\big(e^{(M-1)(\overline v-\underline v_M^{U})/(\overline v-\pi)}-1\big)-(\overline v-\underline v_M^{U}),
\]
which simplifies to $e^{(M-1)(\overline v-\underline v_M^{U})/(\overline v-\pi)}=M$, i.e.,
\begin{equation}\label{eq:vMU-closedform}
\underline v_M^{U}\;=\;\overline v-(\overline v-\pi)\,\frac{\log M}{M-1}.
\end{equation}
Substituting \eqref{eq:vMU-closedform} into the closed form \eqref{eq:aggvolclosedform} applied to $F_U$,
\begin{equation}\label{eq:AU-closedform}
L\,M\int_{\underline v_M^{U}}^{\overline v}\tilde{Q}(x)\,dF_U(x)
\;=\;
L\,\frac{\overline v-\pi}{2}\cdot\frac{M\big(M-1-\log M\big)}{(M-1)^2}.
\end{equation}
The standard inequality $\log M>1-1/M$ for $M>1$ is equivalent to $M(M-1-\log M)<(M-1)^2$. Hence, for all $M>1$,
\begin{equation}\label{eq:AU-below-L}
L\,M\int_{\underline v_M^{U}}^{\overline v}\tilde{Q}(x)\,dF_U(x)
\;<\;
L\,\frac{\overline v-\pi}{2}.
\end{equation}

\medskip
\noindent
Next, we establish tail dominance of $F$ over $F_U$. Assume $M$ is large enough that $(\overline v-\pi)\log M/(M-1)<\eta$, equivalently $\underline v_M^{U}>\overline v-\eta$ by \eqref{eq:vMU-closedform}. For any $u\in[\underline v_M^{U},\overline v]\subset[\overline v-\eta,\overline v]$, integrating the hypothesis $f(u)\le 1/(\overline v-\pi)$ yields
\begin{equation}\label{eq:tail-dom}
1-F(u)\;=\;\int_u^{\overline v}f(s)\,ds\;\le\;\frac{\overline v-u}{\overline v-\pi}\;=\;1-F_U(u).
\end{equation}

\medskip
\noindent
We now compare the cutoffs $\underline v_M$ and $\underline v_M^{U}$. For any continuous CDF $G$ on $[\pi,\overline v]$ and $v\in[\pi,\overline v]$, define
\[
\mathcal I(v;G)\;:=\;\int_v^{\overline v}\!\big(e^{(M-1)(1-G(u))}-1\big)\,du.
\]
By \eqref{eq:cutoff}, $\underline v_M$ is the unique root in $[\pi,\overline v)$ of $v\mapsto \mathcal I(v;F)-(v-\pi)$. This map is strictly decreasing, since its derivative equals $-e^{(M-1)(1-F(v))}<0$. Evaluating at $v=\underline v_M^{U}$ and using \eqref{eq:tail-dom} together with the monotonicity of $x\mapsto e^{(M-1)x}$,
\[
\mathcal I(\underline v_M^{U};F)
\;\le\;
\mathcal I(\underline v_M^{U};F_U)
\;=\;
\underline v_M^{U}-\pi,
\]
where the last equality is \eqref{eq:cutoff} applied to $F_U$ at its own cutoff. Thus $\mathcal I(\underline v_M^{U};F)-(\underline v_M^{U}-\pi)\le 0$, and the strict monotonicity yields
\begin{equation}\label{eq:cutoff-comp}
\underline v_M\;\le\;\underline v_M^{U}.
\end{equation}

\medskip
\noindent
We now conclude. Using the closed form \eqref{eq:aggvolclosedform}, \eqref{eq:cutoff-comp}, and \eqref{eq:AU-below-L},
\[
L\,M\int_{\underline v_M}^{\overline v}\tilde{Q}(x)\,dF(x)
\;=\;
\frac{L\,M\,(\underline v_M-\pi)}{2\,(M-1)}
\;\le\;
\frac{L\,M\,(\underline v_M^{U}-\pi)}{2\,(M-1)}
\;<\;
L\,\frac{\overline v-\pi}{2},
\]
which establishes \eqref{eq:aggvolsaturation} for every $M$ satisfying $(\overline v-\pi)\log M/(M-1)<\eta$.

\medskip
\noindent
We finally establish convergence of the expected end-of-block price. Each unit of volume executed in the block moves the DEX price by $2/L$; see \eqref{eq:impact}. Since the initial price is normalized to zero, the expected end-of-block price is
\[
\frac{2}{L}\,M\int_{\underline v_M}^{\overline v}Q(v)\,dF(v)
\;=\;
2\,M\int_{\underline v_M}^{\overline v}\tilde{Q}(v)\,dF(v)
\;=\;
\frac{M(\underline v_M-\pi)}{M-1},
\]
where the last equality uses the closed form \eqref{eq:aggvolclosedform}. By Lemma~\ref{lem:cutoff}, $\underline v_M\to\overline v$, so the expected end-of-block price converges to $\overline v-\pi$ as $M\to\infty$. Since the expected end-of-block price equals $2/L$ times the expected aggregate trading volume, \eqref{eq:aggvolsaturation} yields $M(\underline v_M-\pi)/(M-1)<\overline v-\pi$ for all $M$ sufficiently large; hence the price limit is also approached from below.

\qed

\subsection{Proof of Proposition \ref{prop:liquidity supply}}

We first derive $\mathbb V[\Delta_M]$. Recall the LP's stage-two prior stated in Section~\ref{sec:solution stage two}: with probability~$1/2$, the liquidation value satisfies $V>0$ and the informed traders' valuations $v_1,\dots,v_M$ are i.i.d.\ from~$F$ on $[0,\overline v]$ (each trader buys); with probability~$1/2$, $V<0$ and valuations are i.i.d.\ from the symmetric density $f(-\cdot)$ on $[-\overline v,0]$ (each trader sells). Equilibrium volumes are $Q(v_k)=L\,\tilde{Q}(v_k)$ on $V>0$ by Proposition~\ref{prop:volume}, and $-Q(|v_k|)$ on $V<0$ by the buy/sell symmetry recorded after Proposition~\ref{prop:eqpfwithvol}.

By construction, the conditional distribution of $Q(v_k)$ on $\{V<0\}$ equals the negative of its conditional distribution on $\{V>0\}$, so
\[
\mathbb E[Q(v_k)]\;=\;\tfrac12\,\mathbb E[Q(v_k)\mid V>0]+\tfrac12\,\mathbb E[Q(v_k)\mid V<0]\;=\;0,
\]
hence $\mathbb E[\Delta_M]=0$ and $\mathbb V[\Delta_M]=\mathbb E[\Delta_M^2]$. Conditional on the sign of $V$, the $v_k$ are i.i.d., so the $Q(v_k)$ are conditionally i.i.d.\ as well. Expanding the square,
\begin{equation}\label{eq:VarDeltaDecomp}
\mathbb E[\Delta_M^2]
\;=\;M\,\mathbb E[Q(v_k)^2]\;+\;M(M-1)\,\mathbb E[Q(v_k)\,Q(v_l)]
\qquad(k\neq l).
\end{equation}

For the diagonal term, using $Q(v_k)=L\,\tilde{Q}(v_k)$ on $\{v_k\ge\underline v_M\}$, $Q(v_k)=0$ on $\{v_k<\underline v_M\}$ (Proposition~\ref{prop:volume}), and the symmetry of the prior across signs,
\begin{equation}\label{eq:DiagonalMoment}
\mathbb E[Q(v_k)^2]
\;=\;\mathbb E[Q(v_k)^2\mid V>0]
\;=\;L^2\int_{\underline v_M}^{\overline v}\tilde{Q}(u)^2\,dF(u).
\end{equation}

For the cross term, with $k\neq l$, conditional independence gives, on each side of the prior,
\[
\mathbb E[Q(v_k)\,Q(v_l)\mid V>0]
\;=\;\bigl(\mathbb E[Q(v_k)\mid V>0]\bigr)^2
\;=\;\Bigl(L\!\int_{\underline v_M}^{\overline v}\tilde{Q}(u)\,dF(u)\Bigr)^{\!2},
\]
and by the symmetric structure $\mathbb E[Q(v_k)\,Q(v_l)\mid V<0]=(-1)^2\bigl(\mathbb E[Q(v_k)\mid V>0]\bigr)^2$ equals the same value. Combining and applying the closed form~\eqref{eq:aggvolclosedform}, $\int_{\underline v_M}^{\overline v}\tilde{Q}(u)\,dF(u)=(\underline v_M-\pi)/(2(M-1))$, yields
\begin{equation}\label{eq:CrossMoment}
\mathbb E[Q(v_k)\,Q(v_l)]
\;=\;L^2\,\frac{(\underline v_M-\pi)^2}{4\,(M-1)^2}.
\end{equation}

Substituting~\eqref{eq:DiagonalMoment}--\eqref{eq:CrossMoment} into~\eqref{eq:VarDeltaDecomp},
\begin{equation}\label{eq:VarDeltaClosedForm}
\mathbb V[\Delta_M]
\;=\;L^2\,M\!\int_{\underline v_M}^{\overline v}\tilde{Q}(u)^2\,dF(u)
\;+\;L^2\,\frac{M\,(\underline v_M-\pi)^2}{4\,(M-1)}
\;=\;L^2\,M\,S_M,
\end{equation}
where the last equality uses the definition~\eqref{eq:defSM}. Combined with $\mathbb E[\Delta_M]=0$, this gives the rightmost equality in~\eqref{eq:loss lp 0}.

Next, we derive the zero-profit closed form. Multiplying the zero-profit condition~\eqref{eq:optimisation problem LP} by $(L^\star+\theta)/L^\star$ (with $L^\star>0$, addressed below) gives
\[
\pi\,N\;=\;(L^\star+\theta)\,M\,S_M,
\]
which rearranges to the closed form~\eqref{eq:liq supp eq}:
\(
L^\star(M)=\pi\,N/(M\,S_M)-\theta.
\)
The supply increases in $\pi\,N$, since $S_M$ is independent of~$\pi\,N$.

We next establish the viability condition. Non-negativity of~\eqref{eq:liq supp eq} requires $\pi\,N/(M\,S_M)\ge\theta$, equivalent to~\eqref{eq:profitability condition}: $M\,S_M\le\pi\,N/\theta$.

We now derive the asymptotic limit. By~\eqref{eq:defSM},
\begin{equation}\label{eq:MSMdecomp}
M\,S_M
\;=\;M\!\int_{\underline v_M}^{\overline v}\tilde{Q}(u)^2\,dF(u)
\;+\;\frac{M\,(\underline v_M-\pi)^2}{4\,(M-1)}.
\end{equation}
The cross-term contribution converges to $(\overline v-\pi)^2/4$ since, by Lemma~\ref{lem:cutoff}, $\underline v_M\to\overline v$ and $M/(M-1)\to 1$. For the integral, we now show
\(
\lim_{M\to\infty}M\!\int_{\underline v_M}^{\overline v}\!\tilde{Q}(u)^2\,dF(u)=(\overline v-\pi)^2/8,
\)
which combined with the cross-term limit gives $\lim_{M\to\infty}M\,S_M=3\,(\overline v-\pi)^2/8$, hence by~\eqref{eq:liq supp eq},
\[
\lim_{M\to\infty}L^\star(M)\;=\;\frac{8\,\pi\,N}{3\,(\overline v-\pi)^2}-\theta,
\]
which is~\eqref{eq:limit kappa}. To distinguish the two contributions to $M\,S_M$, denote
\begin{equation}\label{eq:defSMold}
\widehat S_M\;:=\;\int_{\underline v_M}^{\overline v}\tilde{Q}(u)^2\,dF(u),
\qquad\text{so that}\qquad
S_M\;=\;\widehat S_M\;+\;\frac{(\underline v_M-\pi)^2}{4\,(M-1)}.
\end{equation}

It remains to show that $\lim_{M\to\infty}M\,\widehat S_M=(\overline v-\pi)^2/8$. Using the factored form of $\tilde{Q}$ from the proof of Proposition~\ref{prop:activetraders},
\[
\tilde{Q}(x) = \frac{1}{2}\,e^{-(M-1)(1-F(x))}\left((\overline v-\pi) - B(x)\right),
\qquad
B(x) \equiv \int_x^{\overline v} e^{(M-1)(1-F(u))}\,du,
\]
so that
\[
M\,\widehat S_M
= \frac{M}{4}\int_{\underline v_M}^{\overline v}
e^{-2(M-1)(1-F(x))}\left((\overline v-\pi)-B(x)\right)^2 dF(x)
= T_1 + T_2 + T_3,
\]
where
\begin{align*}
T_1 &= \frac{M(\overline v-\pi)^2}{4}
       \int_{\underline v_M}^{\overline v} e^{-2(M-1)(1-F(x))}\,dF(x),\\
T_2 &= -\frac{M(\overline v-\pi)}{2}
       \int_{\underline v_M}^{\overline v} e^{-2(M-1)(1-F(x))}\,B(x)\,dF(x),\\
T_3 &= \frac{M}{4}
       \int_{\underline v_M}^{\overline v} e^{-2(M-1)(1-F(x))}\,B(x)^2\,dF(x).
\end{align*}

For $T_1$, the change of variable $z=1-F(x)$ gives
\[
T_1
= \frac{M(\overline v-\pi)^2}{4}
  \int_0^{1-F(\underline v_M)} e^{-2(M-1)z}\,dz
= \frac{M(\overline v-\pi)^2}{8(M-1)}
  \left(1 - e^{-2(M-1)(1-F(\underline v_M))}\right).
\]
Since $(M-1)(1-F(\underline v_M))\to\infty$ (established in the proof of
Proposition~\ref{prop:activetraders}), the exponential vanishes and
$M/(M-1)\to 1$, so $T_1\to\frac{(\overline v-\pi)^2}{8}$.

For $T_2$, reversing the order of integration between $x$ and $u$,
\[
\int_{\underline v_M}^{\overline v} e^{-2(M-1)(1-F(x))}B(x)\,dF(x)
= \int_{\underline v_M}^{\overline v} e^{(M-1)(1-F(u))}
  \left[\int_{\underline v_M}^{u} e^{-2(M-1)(1-F(x))}\,dF(x)\right]du.
\]
The inner integral evaluates, via $z=1-F(x)$, to
\[
\int_{\underline v_M}^{u} e^{-2(M-1)(1-F(x))}\,dF(x)
= \frac{1}{2(M-1)}
  \left(e^{-2(M-1)(1-F(u))} - e^{-2(M-1)(1-F(\underline v_M))}\right).
\]
Substituting back and using the cutoff definition \eqref{eq:cutoff},
which gives $\int_{\underline v_M}^{\overline v}e^{(M-1)(1-F(u))}\,du = \overline v-\pi$,
\[
T_2
= -\frac{M(\overline v-\pi)}{4(M-1)}
  \left[
    \int_{\underline v_M}^{\overline v} e^{-(M-1)(1-F(u))}\,du
    \;-\;
    (\overline v-\pi)\,e^{-2(M-1)(1-F(\underline v_M))}
  \right].
\]
The first term in brackets is bounded by $\overline v - \underline v_M \to 0$; the
second vanishes since $(M-1)(1-F(\underline v_M))\to\infty$. Hence $T_2\to 0$.

For $T_3$, since $B$ is decreasing in $x$, we have
$B(x)\le B(\underline v_M)=\overline v-\pi$ for all $x\in[\underline v_M,\overline v]$,
so
\[
0 \le T_3
\le \frac{(\overline v-\pi)}{4}\cdot
     M\int_{\underline v_M}^{\overline v} e^{-2(M-1)(1-F(x))}B(x)\,dF(x)
\to 0,
\]
where the last limit follows from the computation for $T_2$.

\medskip\noindent
Combining the three limits gives
$\lim_{M\to\infty}M\,\widehat S_M = \frac{(\overline v-\pi)^2}{8}$,
which together with the cross-term limit $\lim_{M\to\infty}M\,(\underline v_M-\pi)^2/(4(M-1))=(\overline v-\pi)^2/4$ established above completes Step~4.

Finally, we establish convergence from above by showing that
\begin{equation}\label{eq:MSsaturation}
M\,S_M \;<\; \frac{3\,(\overline v-\pi)^2}{8}
\end{equation}
for all $M$ sufficiently large, which by~\eqref{eq:liq supp eq} is equivalent to $L^\star(M)>\lim_{M\to\infty}L^\star(M)$. By~\eqref{eq:MSMdecomp}, it suffices to prove the two strict inequalities
\begin{align}
M\,\widehat S_M &\;<\;\frac{(\overline v-\pi)^2}{8},\label{eq:MSwidehatStrict}\\
\frac{M\,(\underline v_M-\pi)^2}{4\,(M-1)} &\;<\;\frac{(\overline v-\pi)^2}{4}.\label{eq:CrossStrict}
\end{align}

\medskip
\noindent
We first prove~\eqref{eq:MSwidehatStrict}. Applying the change of variable $z=1-F(x)$ to the expression for $M\,\widehat S_M$ recalled above,
\begin{equation}\label{eq:MSz}
M\,\widehat S_M \;=\; \frac{M}{4}\int_0^{1-F(\underline v_M)}\!\!\!\!e^{-2(M-1) z}\bigl((\overline v-\pi)-B(x)\bigr)^2\,dz,
\end{equation}
with $x=F^{-1}(1-z)$; recall $B(\overline v)=0$ and, by \eqref{eq:cutoff}, $B(\underline v_M)=\overline v-\pi$. By the standing assumption, there exists $\eta>0$ such that $f\le 1/(\overline v-\pi)$ on $[\overline v-\eta,\overline v]$; since $\underline v_M\to\overline v$ (proof of Proposition~\ref{prop:activetraders}), we assume $M$ large enough that $\underline v_M\ge\overline v-\eta$, so that $1/f(u)\ge \overline v-\pi$ for all $u\in[\underline v_M,\overline v]$.

\medskip
\noindent
We first derive a lower bound on $B$. For $z\in[0,1-F(\underline v_M)]$ and $x=F^{-1}(1-z)$, the change of variable $s=1-F(u)$ in $B(x)=\int_x^{\overline v}e^{(M-1)(1-F(u))}\,du$ gives
\[
B(x)\;=\;\int_0^z \frac{e^{(M-1) s}}{f(u)}\,ds
\;\ge\;(\overline v-\pi)\int_0^z e^{(M-1) s}\,ds
\;=\;\frac{\overline v-\pi}{M-1}\bigl(e^{(M-1) z}-1\bigr),
\]
where the inequality uses $1/f(u)\ge \overline v-\pi$ on $[\underline v_M,\overline v]$, the range over which $u$ varies as $s$ ranges over $[0,z]$.

\medskip
\noindent
Evaluating the lower bound just derived at $z=1-F(\underline v_M)$ (so $x=\underline v_M$) and using $B(\underline v_M)=\overline v-\pi$ yields $\overline v-\pi\ge (\overline v-\pi)\bigl(e^{(M-1)(1-F(\underline v_M))}-1\bigr)/(M-1)$, which rearranges to
\begin{equation}\label{eq:betaupper}
1-F(\underline v_M)\;\le\;\frac{\log M}{M-1}.
\end{equation}

\medskip
\noindent
Rearranging the lower bound,
\[
(\overline v-\pi)-B(x)\;\le\;(\overline v-\pi)-\frac{\overline v-\pi}{M-1}\bigl(e^{(M-1) z}-1\bigr)\;=\;\frac{(\overline v-\pi)\bigl(M-e^{(M-1) z}\bigr)}{M-1}.
\]
The left-hand side is nonnegative because $B(x)\le B(\underline v_M)=\overline v-\pi$ (as $B$ is decreasing in $x$), and the right-hand side is nonnegative by \eqref{eq:betaupper}. Squaring preserves the inequality:
\begin{equation}\label{eq:squaredbound}
\bigl((\overline v-\pi)-B(x)\bigr)^2\;\le\;\frac{(\overline v-\pi)^2}{(M-1)^2}\bigl(M-e^{(M-1) z}\bigr)^2\qquad\text{on }[0,1-F(\underline v_M)].
\end{equation}

\medskip
\noindent
Inserting \eqref{eq:squaredbound} into \eqref{eq:MSz} and extending the integration range from $[0,1-F(\underline v_M)]$ to $[0,\log M/(M-1)]$ (valid by \eqref{eq:betaupper}, since the integrand is nonnegative),
\[
M\,\widehat S_M\;\le\;\frac{M(\overline v-\pi)^2}{4(M-1)^2}\int_0^{\log M/(M-1)}\!\!\!\! e^{-2(M-1) z}\bigl(M-e^{(M-1) z}\bigr)^2\,dz.
\]
Direct evaluation of the integral gives
\begin{equation}\label{eq:MSclosed}
M\,\widehat S_M\;\le\;\frac{(\overline v-\pi)^2}{8}\left[\frac{M(M-3)}{(M-1)^2}+\frac{2\,M\,\log M}{(M-1)^3}\right].
\end{equation}

\medskip
\noindent
It remains to verify the strict inequality. Setting $\nu:=M-1\ge 1$ and $g(\nu):=\nu(\nu+2)-2(\nu+1)\log(\nu+1)$, the bracketed expression in \eqref{eq:MSclosed} is strictly less than $1$ if and only if $g(\nu)>0$. One has $g'(\nu)=2\nu-2\log(\nu+1)$ with $g'(0)=0$, and $g''(\nu)=2\nu/(\nu+1)\ge 0$; hence $g$ is non-decreasing on $[0,\infty)$. Since $g(1)=3-4\log 2>0$, monotonicity yields $g(\nu)\ge g(1)>0$ for all $\nu\ge 1$. Combined with \eqref{eq:MSclosed}, this establishes \eqref{eq:MSwidehatStrict} for every $M$ with $\underline v_M\ge\overline v-\eta$.

\medskip
\noindent
We now prove~\eqref{eq:CrossStrict}. Working under the same standing assumption $\underline v_M\ge\overline v-\eta$, we use~\eqref{eq:betaupper} together with the cutoff identity~\eqref{eq:cutoff} to bound $\overline v-\underline v_M$ from below. For $u\in[\underline v_M,\overline v]$, $F(u)\ge F(\underline v_M)$, hence $(M-1)(1-F(u))\le(M-1)(1-F(\underline v_M))\le\log M$ by~\eqref{eq:betaupper}, so $e^{(M-1)(1-F(u))}-1\le M-1$. Substituting into~\eqref{eq:cutoff},
\[
\underline v_M-\pi
\;=\;\int_{\underline v_M}^{\overline v}\bigl(e^{(M-1)(1-F(u))}-1\bigr)\,du
\;\le\;(M-1)\,(\overline v-\underline v_M),
\]
which rearranges to
\begin{equation}\label{eq:cutoffLowerGap}
\overline v-\underline v_M\;\ge\;\frac{\underline v_M-\pi}{M-1}.
\end{equation}
Setting $\gamma:=(\underline v_M-\pi)/(\overline v-\pi)\in(0,1]$, inequality~\eqref{eq:cutoffLowerGap} divided by $\overline v-\pi$ reads $1-\gamma\ge\gamma/(M-1)$, equivalently $(M-1)(1-\gamma)\ge\gamma$, i.e., $M\ge M\gamma+1\ge M\gamma$. Hence $\gamma\le(M-1)/M<1$, so
\[
\frac{M\,(\underline v_M-\pi)^2}{4\,(M-1)}
\;=\;\frac{M\,\gamma^2\,(\overline v-\pi)^2}{4\,(M-1)}
\;\le\;\frac{M\,(M-1)^2/M^2\,(\overline v-\pi)^2}{4\,(M-1)}
\;=\;\frac{(M-1)\,(\overline v-\pi)^2}{4\,M}
\;<\;\frac{(\overline v-\pi)^2}{4},
\]
which is~\eqref{eq:CrossStrict}. Combining~\eqref{eq:MSwidehatStrict} and~\eqref{eq:CrossStrict} via~\eqref{eq:MSMdecomp} yields~\eqref{eq:MSsaturation} for every $M$ with $\underline v_M\ge\overline v-\eta$, hence $L^\star(M)$ is approached from above as $M\to\infty$.\qed

\subsection{Proof of Proposition \ref{prop:eq nb traders}}
Use the expected wealth of trader $i$, conditional on observing a signal $v_i$, in \eqref{eq:expwealth},
\[
\mathbb{E}_{i}\!\big[W_{i}\big]
=
-\Phi_{i}
+Q_{i}\Big(v_{i}-\pi-\frac{Q_{i}}{L}\Big)
-C
-\frac{2\,Q_{i}}{L}
\sum_{j=0}^{M-1}
\mathbb{E}_{i}\!\left[
\mathbbm{1}_{\{\Phi_{(j)}<\Phi_{i}<\Phi_{(j+1)}\}}
\,\Delta_{(j+1:M-1)}
\right],
\]
together with the equilibrium priority fees (Proposition \ref{prop:eqpfwithvol}), trading volumes (Proposition \ref{prop:volume}), and liquidity (Proposition \ref{prop:liquidity supply}), to obtain for each $i\in\{1,\dots,M\}$:
\begin{align*}
\mathbb{E}\big[W_{i}\mid v_{i}\big]
=&
-\frac{2}{L^{\star}}\,(M-1)
\left(
\int_{0}^{Q^{\star}(v_{i})} x^{2}\,dG(x)
+Q^{\star}(v_{i})\int_{Q^{\star}(v_{i})}^{\overline{Q}} x\,dG(x)
\right)
\\
&\quad
+Q^{\star}(v_{i})
\left(
v_{i}-\pi-\frac{Q^{\star}(v_{i})}{L^{\star}}
\right)
-C .
\end{align*}
Using that $Q^{\star}(v_i)=0$ for $v_i<\underline v_M$ and
$Q^{\star}(v_i)=L^{\star}\tilde{Q}(v_i)$ for $v_i\ge\underline v_M$, this can be written as
\begin{align*}
\mathbb{E}\big[W_{i}\mid v_{i}\big]
=
L^{\star}
\Bigg[
&
\tilde{Q}(v_{i})
\big(v_{i}-\pi-\tilde{Q}(v_{i})\big)
\\
&\qquad
-2\,(M-1)
\left(
\int_{0}^{v_{i}}\tilde{Q}(u)^{2}\,dF(u)
+\tilde{Q}(v_{i})
\int_{v_{i}}^{\overline{v}}\tilde{Q}(u)\,dF(u)
\right)
\Bigg]
-C .
\end{align*}

\medskip
\noindent
At stage one, signals have not yet been realized and traders are ex ante symmetric. If $v_i<\underline v_M$, then $Q^{\star}(v_i)=0$ and $\Phi_i=0$, so the payoff is $-C$. Combining with the case $v_i\ge\underline v_M$, the expected wealth is
\begin{align*}
\mathbb{E}\!\left[W_{i}\right]
=&
-C
+L^{\star}
\int_{\underline{v}_{M}}^{\overline{v}}
\tilde{Q}(v)
\big(v-\pi-\tilde{Q}(v)\big)
\,dF(v)
\\
&\quad
-2\,L^{\star}(M-1)
\int_{\underline{v}_{M}}^{\overline{v}}
\left[
\int_{\underline{v}_{M}}^{v}\tilde{Q}(u)^{2}\,dF(u)
+\tilde{Q}(v)\int_{v}^{\overline{v}}\tilde{Q}(u)\,dF(u)
\right]
dF(v).
\end{align*}

\medskip
\noindent
We now show that $\mathbb E[W_i]$ converges to a finite negative limit as $M\to\infty$. First, by Proposition \ref{prop:liquidity supply}, $L^{\star}$ converges to a finite constant as $M\to\infty$.

\medskip
\noindent
Consider
\[
(M-1)
\int_{\underline{v}_{M}}^{\overline{v}}
\int_{\underline{v}_{M}}^{v}
\tilde{Q}(u)^{2}\,dF(u)\,dF(v).
\]
Interchanging the order of integration yields
\begin{align*}
\int_{\underline{v}_{M}}^{\overline{v}}
\int_{\underline{v}_{M}}^{v}
\tilde{Q}(u)^{2}\,dF(u)\,dF(v)
&=
\int_{\underline{v}_{M}}^{\overline{v}}
\int_{u}^{\overline v}
\tilde{Q}(u)^{2}\,dF(v)\,dF(u)
\\
&=
\int_{\underline{v}_{M}}^{\overline{v}}
\tilde{Q}(u)^{2}\big(1-F(u)\big)\,dF(u).
\end{align*}
Since $1-F(u)\le 1$, the limit derived in the proof of Proposition \ref{prop:liquidity supply} implies
\[
0
\le
\limsup_{M\rightarrow\infty}
(M-1)
\int_{\underline{v}_{M}}^{\overline{v}}
\int_{\underline{v}_{M}}^{v}
\tilde{Q}(u)^{2}\,dF(u)\,dF(v)
\le
\frac{(\overline v-\pi)^2}{8}.
\]

\medskip
\noindent
Next, consider
\[
(M-1)
\int_{\underline{v}_{M}}^{\overline{v}}
\int_{v}^{\overline{v}}
\tilde{Q}(v)\tilde{Q}(u)\,dF(u)\,dF(v).
\]
Using the bound $\tilde{Q}(u)\le\tilde{Q}(\overline v)=\frac{\overline v-\pi}{2}$,
\begin{align*}
\int_{\underline{v}_{M}}^{\overline{v}}
\int_{v}^{\overline{v}}
\tilde{Q}(v)\tilde{Q}(u)\,dF(u)\,dF(v)
&\le
\frac{\overline v-\pi}{2}
\int_{\underline{v}_{M}}^{\overline{v}}
\tilde{Q}(v)\,(1-F(v))\,dF(v)
\\
&\le
\frac{\overline v-\pi}{2}
\int_{\underline{v}_{M}}^{\overline{v}}
\tilde{Q}(v)\,dF(v).
\end{align*}
By the limit derived in Proposition \ref{prop:activetraders},
\[
0
\le
\limsup_{M\rightarrow\infty}
(M-1)
\int_{\underline{v}_{M}}^{\overline{v}}
\int_{v}^{\overline{v}}
\tilde{Q}(v)\tilde{Q}(u)\,dF(u)\,dF(v)
\le
\frac{(\overline v-\pi)^2}{4}.
\]

\medskip
\noindent
Collecting terms, the indifference condition $\mathbb E[W_i]=0$ rearranges to
\begin{align*}
C=&
L^{\star}
\int_{\underline{v}_{M}}^{\overline{v}}
\tilde{Q}(v)
\big(v-\pi-\tilde{Q}(v)\big)
\,dF(v)
\\
&\quad
-2\,L^{\star}(M-1)
\int_{\underline{v}_{M}}^{\overline{v}}
(1-F(u))\tilde{Q}(u)^{2}\,dF(u)
\\
&\quad
-2\,L^{\star}(M-1)
\int_{\underline{v}_{M}}^{\overline{v}}
\int_{v}^{\overline{v}}
\tilde{Q}(v)\tilde{Q}(u)\,dF(u)\,dF(v).
\end{align*}
The two $(M-1)$ terms remain bounded as $M\to\infty$, while the first integral term converges to zero. Hence,
\[
\limsup_{M\rightarrow\infty}\mathbb{E}\!\left[W_{i}\right] < 0.
\]
Therefore, if $\mathbb{E}[W_i]>0$ for $M=2$, there exists a finite equilibrium number of informed traders, defined as the largest integer $M$ such that $\mathbb{E}[W_i]\ge 0$.

\medskip
\noindent
\emph{Simplification of $H(M)$.} The volume FOC \eqref{eq:integralEqVolume2}, after substituting $Q=L^\star\tilde Q$ and multiplying by $\tilde Q(v)$, yields the identity
\[
\tilde Q(v)\bigl(v-\pi-\tilde Q(v)\bigr) \;=\; \tilde Q(v)^2 \;+\; 2(M-1)\,\tilde Q(v)\!\int_v^{\overline v}\!\tilde Q(u)\,dF(u),\qquad v\in[\underline v_M,\overline v].
\]
Substituting into the first integral of the indifference condition above, the $\tilde Q(v)\tilde Q(u)$ cross term cancels exactly with the third integral, leaving
\[
C \;=\; L^\star\!\int_{\underline v_M}^{\overline v}\!\tilde Q(v)^2\bigl[1-2(M-1)(1-F(v))\bigr]\,dF(v) \;=\; H(M),
\]
the form in~\eqref{eq:eqconditionM}.

\noindent We now establish the comparative statics. Decreasing $C$ weakly enlarges the set $\{M\ge 2 : C\le H(M)\}$, so $M^\star$ is weakly decreasing in $C$. By Proposition~\ref{prop:liquidity supply}, $L^\star(M)$ is increasing in $N$ (with $\pi$ fixed), while the bracketed integrand in~\eqref{eq:eqconditionM} does not depend on $N$. At any $M$ with $H(M)\ge C\ge 0$ and $L^\star(M)>0$, the integral over that bracket is nonnegative, so raising $N$ weakly raises $H(M)$ and thus weakly raises $M^\star$. \qed

\subsection{Proof of Corollary~\ref{cor:blocktime}}

Throughout, $F(\cdot\,;T)$ denotes the distribution of private valuations with support $[0,\overline v(T)]$ and $N(T)$ the mass of noise-trading flow.

We first show that the participation cutoff $\underline v_M(T)$ is strictly increasing in $T$. Define
\[
G(v;T)\;:=\;v-\pi-\int_v^{\overline v(T)}\!\Bigl(e^{(1-F(u;T))(M-1)}-1\Bigr)\,du,
\]
so that the cutoff $\underline v_M(T)$ satisfies $G(\underline v_M(T);T)=0$ by~\eqref{eq:cutoff}. Differentiating in $v$,
\[
\partial_v G(v;T)\;=\;1+\Bigl(e^{(1-F(v;T))(M-1)}-1\Bigr)\;=\;e^{(1-F(v;T))(M-1)}\;>\;0.
\] 
For $T'>T$, the FOSD ordering in Assumption~\ref{assume:T}(i) yields $F(u;T')<F(u;T)$ on $[0,\overline v(T)]$, so
\[
e^{(1-F(u;T'))(M-1)}-1\;>\;e^{(1-F(u;T))(M-1)}-1\qquad\text{for }u\in{(0,\overline v(T)]}.
\]
Moreover, on $(\overline v(T),\overline v(T')]$ one has $F(u;T')<1$, hence $e^{(1-F(u;T'))(M-1)}-1>0$. Combining,
\[
\int_v^{\overline v(T')}\!\Bigl(e^{(1-F(u;T'))(M-1)}-1\Bigr)\,du\;>\;\int_v^{\overline v(T)}\!\Bigl(e^{(1-F(u;T))(M-1)}-1\Bigr)\,du,
\]
so $G(v;T')<G(v;T)$ for every $v\in[0,\overline v(T)]$. Since $\partial_v G>0$, the unique solution of $G(\cdot;T')=0$ satisfies $\underline v_M(T')>\underline v_M(T)$.
  
We now show that the end-of-block price is strictly increasing in $T$ in the regime with many informed traders. From Proposition~\ref{prop:activetraders}, as $M\to\infty$ the end-of-block price converges to $\overline v-\pi$. With $\overline v=\overline v(T)$, the limit equals $\overline v(T)-\pi$, which is strictly increasing in $T$ by Assumption~\ref{assume:T}(i).

Next, we show that the equilibrium liquidity $L^\infty(T)$ is decreasing in $T$. From~\eqref{eq:limit kappa} in Proposition~\ref{prop:liquidity supply},
\[
L^\infty(T)\;=\;\frac{8\,\pi\, N(T)}{3\,(\overline v(T)-\pi)^2}-\theta.
\]
By Assumption~\ref{assume:T}(i), $\overline v(T)-\pi$ is strictly increasing in $T$, so the denominator $3\,(\overline v(T)-\pi)^2$ is strictly increasing; by Assumption~\ref{assume:T}(ii), the numerator $8\,\pi\,N(T)$ is decreasing in $T$. Hence $L^\infty(T)$ is strictly decreasing in $T$.

Finally, we show that markets shut down for sufficiently large $T$. Since $N(T)\ge 0$ is decreasing, $8\,\pi\,N(T)\le 8\,\pi\,N(0)$ is bounded. Since $\overline v(T)-\pi\to\infty$ by Assumption~\ref{assume:T}(i), $3\,(\overline v(T)-\pi)^2\to\infty$, so the first term in $L^\infty(T)$ tends to zero, hence $L^\infty(T)\to-\theta<0$. Hence there exists $\overline T<\infty$ such that $L^\infty(T)\le 0$ for all $T\ge\overline T$.\qed

\begin{doublespacing}
\bibliographystyle{jf}
\bibliography{references}

@article{klein2023price,
  title={Informed Liquidity Provision on Decentralized Exchanges},
  author={Klein, Olga and Kozhan, Roman and Viswanath-Natraj, Ganesh and Wang, Junxuan},
  journal={Available at SSRN 4642411},
  year={2023}
}

@article{wadhwa2025aucil,
  title={AUCIL: An Inclusion List Design for Rational Parties},
  author={Wadhwa, Sarisht and Ma, Julian and Thiery, Thomas and Monnot, Barnabe and Zanolini, Luca and Zhang, Fan and Nayak, Kartik},
  journal={Cryptology ePrint Archive},
  year={2025}
}

@misc{thiery2024focil,
  author       = {Thiery, Thomas and D'Amato, Francesco and Ma, Julian and
                  Monnot, Barnab{\'e} and Tsao, Terence and Kaufmann, Jacob and
                  Song, Jihoon},
  title        = {{EIP-7805}: Fork-choice enforced Inclusion Lists ({FOCIL})},
  howpublished = {Ethereum Improvement Proposals, no. 7805},
  year         = {2024},
  month        = {November},
  note         = {Draft. Available: \url{https://eips.ethereum.org/EIPS/eip-7805}},
}

@article{capponi2023price,
  title={Price discovery on decentralized exchanges},
  author={Capponi, Agostino and Jia, Ruizhe and Yu, Shihao},
  journal={Available at SSRN 4236993},
  year={2023}
}

@article{drissi2025equilibrium,
  title={Equilibrium Liquidity and Risk Offsetting in Decentralised Markets},
  author={Drissi, Fay{\c{c}}al and Wu, Xuchen and Jaimungal, Sebastian},
  journal={arXiv preprint arXiv:2512.19838},
  year={2025}
}

@article{lehar2025decentralized,
  title={Decentralized exchange: The uniswap automated market maker},
  author={Lehar, Alfred and Parlour, Christine},
  journal={The Journal of Finance},
  volume={80},
  number={1},
  pages={321--374},
  year={2025},
  publisher={Wiley Online Library}
}

@article{capponi2023decentralized,
  title={Decentralized finance: Protocols, risks, and governance},
  author={Capponi, Agostino and Iyengar, Garud and Sethuraman, Jay and others},
  journal={Foundations and Trends{\textregistered} in Privacy and Security},
  volume={5},
  number={3},
  pages={144--188},
  year={2023},
  publisher={Now Publishers, Inc.}
}

@article{capponi2024maximal,
  title={Maximal extractable value and allocative inefficiencies in public blockchains},
  author={Capponi, Agostino},
  journal={Available at SSRN 4931619},
  year={2024}
}

@article{john2025economics,
  title={Economics of ethereum},
  author={John, Kose and Monnot, Barnab{\'e} and Mueller, Peter and Saleh, Fahad and Schwarz-Schilling, Caspar},
  journal={Journal of Corporate Finance},
  volume={91},
  pages={102718},
  year={2025},
  publisher={Elsevier}
}

@article{capponi2024price,
  title={Price discovery on decentralized exchanges},
  author={Capponi, Agostino and Jia, Ruizhe and Yu, Shihao},
  journal={Available at SSRN 4236993},
  year={2024}
}

@article{biais2023advances,
  title={Advances in blockchain and crypto economics},
  author={Biais, Bruno and Capponi, Agostino and Cong, Lin William and Gaur, Vishal and Giesecke, Kay},
  journal={Management Science},
  volume={69},
  number={11},
  pages={6417--6426},
  year={2023},
  publisher={INFORMS}
}

@article{biais1993price,
  title={Price formation and equilibrium liquidity in fragmented and centralized markets},
  author={Biais, Bruno},
  journal={The Journal of Finance},
  volume={48},
  number={1},
  pages={157--185},
  year={1993},
  publisher={Wiley Online Library}
}

@inproceedings{liu2022empirical,
  title={Empirical analysis of eip-1559: Transaction fees, waiting times, and consensus security},
  author={Liu, Yulin and Lu, Yuxuan and Nayak, Kartik and Zhang, Fan and Zhang, Luyao and Zhao, Yinhong},
  booktitle={Proceedings of the 2022 ACM SIGSAC Conference on Computer and Communications Security},
  pages={2099--2113},
  year={2022}
}

@article{mizrach2025marginal,
  title={The Marginal Effects of Ethereum Network MEV Transaction Re-Ordering},
  author={Mizrach, Bruce and Yoshida, Nathaniel},
  journal={arXiv preprint arXiv:2508.04003},
  year={2025}
}

@article{cartea2024decentralized,
  title={Decentralized Finance and Automated Market Making: Predictable Loss and Optimal Liquidity Provision},
  author={Cartea, {\'A}lvaro and Drissi, Fay{\c{c}}al and Monga, Marcello},
  journal={SIAM Journal on Financial Mathematics},
  volume={15},
  number={3},
  pages={931--959},
  year={2024},
  publisher={SIAM}
}

@article{biais1999price,
  title={Price discovery and learning during the preopening period in the Paris Bourse},
  author={Biais, Bruno and Hillion, Pierre and Spatt, Chester},
  journal={Journal of Political Economy},
  volume={107},
  number={6},
  pages={1218--1248},
  year={1999},
  publisher={The University of Chicago Press}
}

@article{medrano2001strategic,
  title={Strategic behavior and price discovery},
  author={Medrano, Luis Angel and Vives, Xavier},
  journal={RAND Journal of Economics},
  pages={221--248},
  year={2001},
  publisher={JSTOR}
}

@article{barbon2021quality,
  title={On the quality of cryptocurrency markets: Centralized versus decentralized exchanges},
  author={Barbon, Andrea and Ranaldo, Angelo},
  journal={arXiv preprint arXiv:2112.07386},
  year={2021}
}

@article{aoyagi2021coexisting,
  title={Coexisting exchange platforms: Limit order books and automated market makers},
  author={Aoyagi, Jun and Ito, Yuki},
  year={2021},
  publisher={Limit Order Books and Automated Market Makers (March 20, 2021)}
}

@article{heines2021tokenization,
  title={The Tokenization of Everything: Towards a Framework for Understanding the Potentials of Tokenized Assets.},
  author={Heines, Roger and Dick, Christian and Pohle, Christian and Jung, Reinhard},
  journal={PACIS},
  volume={40},
  year={2021}
}

@article{cartea2024strategic,
  title={Strategic bonding curves in automated market makers},
  author={Cartea, {\'A}lvaro and Drissi, Fay{\c{c}}al and S{\'a}nchez-Betancourt, Leandro and Siska, David and Szpruch, Lukasz},
  journal={Available at SSRN 5018420},
  year={2024}
}

@article{cardozo2024cross,
  title={On Cross-Border Crypto Flows},
  author={Cardozo, Pamela and Fern{\'a}ndez, Andr{\'e}s and Jiang, Jerzy and Rojas, Felipe},
  year={2024},
  publisher={IMF Working Paper}
}

@misc{hub2023project,
  title={Project Mariana: Cross-border exchange of wholesale CBDCs using automated market-makers},
  author={Hub, BIS Innovation},
  year={2023}
}

@techreport{fung2016central,
  title={Central bank digital currencies: a framework for assessing why and how},
  author={Fung, Ben Siu-cheong and Halaburda, Hanna},
  year={2016},
  institution={Bank of Canada Staff Discussion Paper}
}

@article{auer2022central,
  title={Central bank digital currencies: motives, economic implications, and the research frontier},
  author={Auer, Raphael and Frost, Jon and Gambacorta, Leonardo and Monnet, Cyril and Rice, Tara and Shin, Hyun Song},
  journal={Annual review of economics},
  volume={14},
  number={1},
  pages={697--721},
  year={2022},
  publisher={Annual Reviews}
}

@article{harvey2016cryptofinance,
  title={Cryptofinance},
  author={Harvey, Campbell R},
  journal={Available at SSRN 2438299},
  year={2016}
}

@book{harvey2021defi,
  title={DeFi and the Future of Finance},
  author={Harvey, Campbell R},
  year={2021},
  publisher={John Wiley \& Sons}
}

@article{cong2019blockchain,
  title={Blockchain disruption and smart contracts},
  author={Cong, Lin William and He, Zhiguo},
  journal={The Review of Financial Studies},
  volume={32},
  number={5},
  pages={1754--1797},
  year={2019},
  publisher={Oxford University Press}
}

@article{john2023smart,
  title={Smart contracts and decentralized finance},
  author={John, Kose and Kogan, Leonid and Saleh, Fahad},
  journal={Annual Review of Financial Economics},
  volume={15},
  number={1},
  pages={523--542},
  year={2023},
  publisher={Annual Reviews}
}

@techreport{van2011transparency,
  title={Transparency and ending times of call auctions: a comparison of Euronext and Xetra},
  author={Van Bommel, Jos and Hoffmann, Peter},
  year={2011},
  institution={Luxembourg School of Finance, University of Luxembourg}
}

@article{hendershott2014price,
  title={Price pressures},
  author={Hendershott, Terrence and Menkveld, Albert J},
  journal={Journal of Financial economics},
  volume={114},
  number={3},
  pages={405--423},
  year={2014},
  publisher={Elsevier}
}

@article{ho1981optimal,
  title={Optimal dealer pricing under transactions and return uncertainty},
  author={Ho, Thomas and Stoll, Hans R},
  journal={Journal of Financial economics},
  volume={9},
  number={1},
  pages={47--73},
  year={1981},
  publisher={Elsevier}
}

@article{holden1992long,
  title={Long-lived private information and imperfect competition},
  author={Holden, Craig W and Subrahmanyam, Avanidhar},
  journal={The Journal of Finance},
  volume={47},
  number={1},
  pages={247--270},
  year={1992},
  publisher={Wiley Online Library}
}

@article{foster1996strategic,
  title={Strategic trading when agents forecast the forecasts of others},
  author={Foster, F Douglas and Viswanathan, S},
  journal={The Journal of Finance},
  volume={51},
  number={4},
  pages={1437--1478},
  year={1996},
  publisher={Wiley Online Library}
}

@article{madhavan1992trading,
  title={Trading mechanisms in securities markets},
  author={Madhavan, Ananth},
  journal={the Journal of Finance},
  volume={47},
  number={2},
  pages={607--641},
  year={1992},
  publisher={Wiley Online Library}
}

@inproceedings{wang2025private,
  title={Private Order Flows and Builder Bidding Dynamics: The Road to Monopoly in Ethereum's Block Building Market},
  author={Wang, Shuzheng and Huang, Yue and Zhang, Wenqin and Huang, Yuming and Wang, Xuechao and Tang, Jing},
  booktitle={Proceedings of the ACM on Web Conference 2025},
  pages={2144--2157},
  year={2025}
}

@article{petryk2025promises,
  title={Promises and Perils of Decentralization in the Blockchain Age},
  author={Petryk, Mariia and M{\"u}ller-Bloch, Christoph and Hahn, Jungpil and Halaburda, Hanna and Henfridsson, Ola and Obermeier, Daniel and Yoo, Youngjin},
  year={2025}
}

@article{harvey2025economic,
  title={An Economic Model of the L1-L2 Interaction},
  author={Harvey, Campbell R and Saleh, Fahad and Sverchkov, Ruslan},
  journal={Available at SSRN 5163823},
  year={2025}
}

@article{baldauf2020high,
  title={High-frequency trading and market performance},
  author={Baldauf, Markus and Mollner, Joshua},
  journal={The Journal of Finance},
  volume={75},
  number={3},
  pages={1495--1526},
  year={2020},
  publisher={Wiley Online Library}
}

@article{lazear1981rank,
  title={Rank-order tournaments as optimum labor contracts},
  author={Lazear, Edward P and Rosen, Sherwin},
  journal={Journal of political Economy},
  volume={89},
  number={5},
  pages={841--864},
  year={1981},
  publisher={The University of Chicago Press}
}

@article{he2025arbitrage,
  title={Arbitrage on decentralized exchanges},
  author={He, Xue Dong and Yang, Chen and Zhou, Yutian},
  journal={arXiv preprint arXiv:2507.08302},
  year={2025}
}

@article{moldovanu2001optimal,
  title={The optimal allocation of prizes in contests},
  author={Moldovanu, Benny and Sela, Aner},
  journal={American Economic Review},
  volume={91},
  number={3},
  pages={542--558},
  year={2001},
  publisher={American Economic Association}
}

@article{hillman1989politically,
  title={Politically contestable rents and transfers},
  author={Hillman, Arye L and Riley, John G},
  journal={Economics \& Politics},
  volume={1},
  number={1},
  pages={17--39},
  year={1989},
  publisher={Wiley Online Library}
}

@article{klose2015all,
  title={The all-pay auction with complete information and identity-dependent externalities},
  author={Klose, Bettina and Kovenock, Dan},
  journal={Economic Theory},
  volume={59},
  number={1},
  pages={1--19},
  year={2015},
  publisher={Springer}
}

@article{barut1998symmetric,
  title={The symmetric multiple prize all-pay auction with complete information},
  author={Barut, Yasar and Kovenock, Dan},
  journal={European Journal of Political Economy},
  volume={14},
  number={4},
  pages={627--644},
  year={1998},
  publisher={Elsevier}
}

@inproceedings{heimbach2024non,
  title={Non-atomic arbitrage in decentralized finance},
  author={Heimbach, Lioba and Pahari, Vabuk and Schertenleib, Eric},
  booktitle={2024 IEEE Symposium on Security and Privacy (SP)},
  pages={3866--3884},
  year={2024},
  organization={IEEE}
}

@article{harvey2024international,
  title={International business and decentralized finance},
  author={Harvey, Campbell R and Rabetti, Daniel},
  journal={Journal of International Business Studies},
  pages={1--24},
  year={2024},
  publisher={Springer}
}

@article{hasbrouck2023economic,
  title={An economic model of a decentralized exchange with concentrated liquidity},
  author={Hasbrouck, Joel and Rivera, Thomas J and Saleh, Fahad},
  journal={Available at SSRN 4529513},
  year={2023}
}

@article{hasbrouck2022need,
  title={The need for fees at a dex: How increases in fees can increase dex trading volume},
  author={Hasbrouck, Joel and Rivera, Thomas J and Saleh, Fahad},
  journal={Available at SSRN 4192925},
  year={2022}
}

@article{biais2023equilibrium,
  title={Equilibrium bitcoin pricing},
  author={Biais, Bruno and Bisiere, Christophe and Bouvard, Matthieu and Casamatta, Catherine and Menkveld, Albert J},
  journal={The Journal of Finance},
  volume={78},
  number={2},
  pages={967--1014},
  year={2023},
  publisher={Wiley Online Library}
}

@article{cong2023scaling,
  title={Scaling smart contracts via layer-2 technologies: Some empirical evidence},
  author={Cong, Lin William and Hui, Xiang and Tucker, Catherine and Zhou, Luofeng},
  journal={Management Science},
  volume={69},
  number={12},
  pages={7306--7316},
  year={2023},
  publisher={INFORMS}
}

@article{cong2021tokenomics,
  title={Tokenomics: Dynamic adoption and valuation},
  author={Cong, Lin William and Li, Ye and Wang, Neng},
  journal={The Review of Financial Studies},
  volume={34},
  number={3},
  pages={1105--1155},
  year={2021},
  publisher={Oxford University Press}
}

@article{ho1983dynamics,
  title={The dynamics of dealer markets under competition},
  author={Ho, Thomas SY and Stoll, Hans R},
  journal={The Journal of Finance},
  volume={38},
  number={4},
  pages={1053--1074},
  year={1983},
  publisher={Wiley Online Library}
}

@article{kyle:1985,
    title = {Continuous Auctions and Insider Trading},
    author = {Kyle, A. S.},
    journal = {Econometrica},
    volume = {53},
    issue = {6},
    pages = {1315--1335}, 
    year = 1985
}

@article{grossman1980impossibility,
  title={On the impossibility of informationally efficient markets},
  author={Grossman, Sanford J and Stiglitz, Joseph E},
  journal={The American economic review},
  volume={70},
  number={3},
  pages={393--408},
  year={1980},
  publisher={JSTOR}
}

@article{kyle1989informed,
  title={Informed speculation with imperfect competition},
  author={Kyle, Albert S},
  journal={The Review of Economic Studies},
  volume={56},
  number={3},
  pages={317--355},
  year={1989},
  publisher={Wiley-Blackwell}
}

@article{john2020proof,
  title={Proof-of-work versus proof-of-stake: A comparative economic analysis},
  author={John, Kose and Rivera, Thomas J and Saleh, Fahad},
  journal={Available at SSRN 3750467},
  year={2020}
}

@article{verrecchia1982information,
  title={Information acquisition in a noisy rational expectations economy},
  author={Verrecchia, Robert E},
  journal={Econometrica: Journal of the Econometric Society},
  pages={1415--1430},
  year={1982},
  publisher={JSTOR}
}

@article{capponi2021adoption,
  title={The adoption of blockchain-based decentralized exchanges},
  author={Capponi, Agostino and Jia, Ruizhe},
  journal={arXiv preprint arXiv:2103.08842},
  year={2021}
}

@inproceedings{vujivcic2018blockchain,
  title={Blockchain technology, bitcoin, and Ethereum: A brief overview},
  author={Vuji{\v{c}}i{\'c}, Dejan and Jagodi{\'c}, Dijana and Ran{\dj}i{\'c}, Sini{\v{s}}a},
  booktitle={2018 17th international symposium infoteh-jahorina (infoteh)},
  pages={1--6},
  year={2018},
  organization={IEEE}
}

@article{angeris2021replicating2,
  title={Replicating Market Makers},
  author={Angeris, Guillermo and Evans, Alex and Chitra, Tarun},
  journal={arXiv preprint arXiv:2103.14769},
  year={2021}
}

@article{milionis2022automated,
  title={Automated market making and loss-versus-rebalancing},
  author={Milionis, Jason and Moallemi, Ciamac C and Roughgarden, Tim and Zhang, Anthony Lee},
  journal={arXiv preprint arXiv:2208.06046},
  year={2022}
}

@article{glosten1985bid,
  title={Bid, ask and transaction prices in a specialist market with heterogeneously informed traders},
  author={Glosten, Lawrence R and Milgrom, Paul R},
  journal={Journal of Financial Economics},
  volume={14},
  number={1},
  pages={71--100},
  year={1985},
  publisher={Elsevier}
}

@article{cartea2025decentralised,
  title={Decentralised finance and automated market making: Execution and speculation},
  author={Cartea, {\'A}lvaro and Drissi, Fay{\c{c}}al and Monga, Marcello},
  journal={Journal of Economic Dynamics and Control},
  pages={105134},
  year={2025},
  publisher={Elsevier}
}

@article{malinova2024learning,
  title={Learning from DeFi: Would automated market makers improve equity trading?},
  author={Malinova, Katya and Park, Andreas},
  journal={Available at SSRN 4531670},
  year={2024}
}

@article{park2023conceptual,
  title={The conceptual flaws of decentralized automated market making},
  author={Park, Andreas},
  journal={Management Science},
  volume={69},
  number={11},
  pages={6731--6751},
  year={2023},
  publisher={INFORMS}
}

@article{cartea2023predictable,
  title={Predictable losses of liquidity provision in constant function markets and concentrated liquidity markets},
  author={Cartea, {\'A}lvaro and Drissi, Fay{\c{c}}al and Monga, Marcello},
  journal={Applied Mathematical Finance},
  volume={30},
  number={2},
  pages={69--93},
  year={2023},
  publisher={Taylor \& Francis}
}

@article{de2002risk,
  title={Risk aversion, transparency, and market performance},
  author={De Frutos, M {\'A}ngeles and Manzano, Carolina},
  journal={The Journal of Finance},
  volume={57},
  number={2},
  pages={959--984},
  year={2002},
  publisher={Wiley Online Library}
}

@article{garman1976market,
  title={Market microstructure},
  author={Garman, Mark B},
  journal={Journal of financial Economics},
  volume={3},
  number={3},
  pages={257--275},
  year={1976},
  publisher={Elsevier}
}

@article{glosten1994electronic,
  title={Is the electronic open limit order book inevitable?},
  author={Glosten, Lawrence R},
  journal={The Journal of Finance},
  volume={49},
  number={4},
  pages={1127--1161},
  year={1994},
  publisher={Wiley Online Library}
}

@article{biais2000competing,
  title={Competing mechanisms in a common value environment},
  author={Biais, Bruno and Martimort, David and Rochet, Jean-Charles},
  journal={Econometrica},
  volume={68},
  number={4},
  pages={799--837},
  year={2000},
  publisher={Wiley Online Library}
}

@book{o1998market,
  title={Market microstructure theory},
  author={O'Hara, Maureen},
  year={1998},
  publisher={John Wiley \& Sons}
}

@article{budish2015high,
  title={The high-frequency trading arms race: Frequent batch auctions as a market design response},
  author={Budish, Eric and Cramton, Peter and Shim, John},
  journal={The Quarterly Journal of Economics},
  volume={130},
  number={4},
  pages={1547--1621},
  year={2015},
  publisher={MIT Press}
}
\end{doublespacing}

\clearpage

\renewcommand{\enotesize}{\normalsize}

\clearpage

\end{doublespace}

\end{document}